# Message-Oriented Middleware Systems: Technology Overview

Wael Al-Manasrah, Zuhair AlSader, Tim Brecht, Ahmed Alquraan, Samer Al-Kiswany

University of Waterloo, Waterloo, Ontario, Canada

*Abstract*—We present a comprehensive characterization study of open-source message-oriented middleware (MOM) systems. We followed a rigorous methodology to select and study ten popular and diverse MOM systems. For each system, we examine 42 features with a total of 134 different options. We found that MOM systems have evolved to provide a framework for modern cloud applications through high flexibility and configurability and by offering core building blocks for complex applications including transaction support, active messaging, resource management, flow control, and native support for multi-tenancy. We also identify that there is an opportunity for the community to consolidate its efforts on fewer open-source projects.

We have also created an annotated data set that makes it easy to verify our findings, which can also be used to help practitioners and developers understand and compare the features of different systems. For a wider impact, we make our data set publicly available.

## I. INTRODUCTION

Modern cloud applications are complex [1]–[3]. They integrate numerous subsystems and they are deployed on hundreds of nodes, increasingly across multiple data centers. Building cloud applications using low-level sockets or synchronous middleware systems such as RPC is cumbersome and lacks critical features required by modern applications, such as reliability, asynchronous delivery, efficient multicast, and load balancing. Consequently, modern applications increasingly use higher-level message-oriented middleware (MOM) systems, also known as messaging, message queuing, or publish/subscribe (pub/sub) systems.

MOM systems are widely used to support a broad range of modern applications, including microservice-based services [4]–[6], streaming analytics [7]–[9], financial services [10], [11], IoT applications [12], video streaming [13]–[15], machine learning services [16]–[18], and blockchain frameworks [19].

At a minimum, MOM systems offer a simple communication abstraction. Producers produce messages to a certain topic hosted at a MOM server (a.k.a. broker). Brokers forward these messages to one or more consumers that have subscribed to that topic (a.k.a. subscribers).

Despite providing a similar communication abstraction, the community maintains tens of open-source MOM systems (Section III), and new systems are frequently introduced. Many of these systems are widely popular, such as Kafka [20], Pulsar [21], RabbitMQ [22], and ActiveMQ Artemis [23]. Many companies offer proprietary MOM systems [24]–[26], and major cloud providers offer MOM services (e.g., Google Cloud Pub/Sub [27], Amazon Simple Queue Service [28], and Microsoft Azure Queue Storage [29]).

This raises a number of questions about the state-of-the-art production MOM systems: What are the differences between the many open-source MOM systems? What are the techniques used to support scalability and reliability? What are the semantics and policies for message delivery, persistence, and ordering? How do MOM systems support flow control and multi-tenancy? Do MOM systems offer a common subset of features? Does an opportunity exist to consolidate the community effort on a smaller subset of systems? Answering these questions is important to guide research and development efforts, improve our understanding of the emerging capabilities of MOM systems and detail the trade-offs offered by different MOM systems, help developers choose the best system for a certain application, help practitioners understand the semantics of different systems and how to configure and maintain these systems appropriately, and help the community understand the capabilities of different systems and focus its effort on a fewer number of MOM systems.

To answer these questions about the state-of-the-art production MOM systems, we conduct a comprehensive study of ten popular MOM systems. For each system, we study 42 features. These features have 134 configuration and deployment options. For each system, we study its documentation, user API, and related posts on developer forums. In a few cases, we checked the source code, contacted the system developers, and deployed the system to verify some characteristics of the system.

We found that MOM systems are evolving and expanding the set of features far beyond simple message communication. First, MOM systems offer a framework for developing complex cloud applications with support for the durable storage of messages, transactions, application-specific processing, multitenancy, resource management, flow control, and configurable message delivery semantics. The majority of the systems we study support complex transactions that can produce messages to and consume messages from multiple topics. Recently, some of these systems have added active messaging capabilities in which applications can submit lambda-style functions to be executed on certain messages, which moves these MOM systems closer to a stream-processing engine or serverless-computing framework.

Second, a striking characteristic of MOM systems is their high flexibility and configurability. Some of the systems offer a flexible topology that allows optimization of the system de-ployment to match the underlying infrastructure or application access patterns. Many of the systems have multiple configuration options for most features we study allowing them to support applications with diverse semantics and performance requirements. We note that multiple systems offer the same features, with Pulsar, RabbitMQ, and ActiveMQ offering the largest number of configurable features. As a result, we believe that it would be beneficial to consolidate the community efforts on fewer projects.

**Reliability**. Most of the systems we study support storing messages on durable storage. The majority of systems support replicating messages for higher reliability and scalability. Messages are stored with strong or eventual consistency semantics, with the majority of systems offering eventually consistent storage semantics. Most systems also incorporate techniques to help consumers recover their state after a failure.

**Service Semantics**. We found that the systems we study adopt four delivery semantics: best-effort, at-most-once, at-least-once, and effectively-once. Interestingly, the majority of the systems support transactions that can consume/produce messages from/to multiple topics. Nevertheless, many systems offer non-atomic transactions. We identified two message-ordering semantics: path ordering and multicast exchange ordering. Multicast exchange ordering guarantees in-order delivery of messages to consumers of a multicast exchange. Path ordering is a weaker model and guarantees that messages are delivered in order only if they take the same path in the system.

**Client Interactions**. All of the systems we study employ a discovery service to provide the information needed for producers and consumers to access the system. The discovery service can either be strongly or eventually consistent. The majority of systems use an eventually consistent discovery service. We identified five approaches to deliver messages to consumers. The majority of systems offer multiple delivery approaches, with pushing messages to consumers being the most common because it incurs the lowest latency.

**Resource Management and Flow Control**. To control the use of available hardware resources, MOM systems can enforce limits on resource usage, including specifying limits on processing, memory, network, and disk resources. The systems we studied also implement flow control mechanisms using two approaches: credit-based and rate-based. These techniques are the base for offering isolation for multi-tenant deployments.

**Active Messaging**. Interestingly, two systems have recently added support for active messaging, in which a user-defined function can consume and produce messages in the system. This capability allows MOM systems to support stream-processing applications and serverless computing with stateful functions. This capability significantly increases system flexibility but introduces new scheduling, isolation, and resource management challenges.

**Protocols**. A number of standards that specify how to interact with MOM systems had been developed. Our study (Section XI) shows that the majority of the systems we study build custom protocols and that the most popular standard, MQTT, is supported by five of the ten systems we study.

We have created a data set in which we annotate each characteristic of each system with a note and a link to the web source detailing the characteristic. We are making our data set publicly available [30] to make it easier to verify our results and for researchers, practitioners, and application developers to find details related to each feature of each system.

The rest of this paper is organized as follows. In Section II, we discuss related work. We present our research methodology in Section III. We present our abstract communication model used by MOM systems in Section IV and discuss their dissemination topologies in Section V. We describe our findings related to reliability in Section VI, service semantics in Section VII, resource management and flow control in Section VIII, active messaging in Section IX, and client interaction in Section X. We present an overview of standard messaging protocols in Section XI. We discuss our findings and concluding remarks in Section XIII. Appendix B presents additional results related to message life cycle in the system and Appendix C presents implementation details related to client interaction.

Throughout the paper, we include notes in grey boxes. These notes add additional technical or deployment information, or give an example related to the section. The reader can skip reading those notes without losing context for the following sections. Furthermore, most of our tables include footnotes that provide additional details about a specific feature in a system. These footnotes are geared toward MOM system experts and can be skipped without any loss of context.

## II. RELATED WORK

We identify two types of related work: surveys that study MOM systems and benchmarking studies that empirically compare MOM systems.

### A. Survey Studies

The work most related to our study is the 20-year-old survey conducted by Eugster et al. [31]. Their work focuses only on pub/sub systems and does not study any message queuing systems. In addition, some of the included systems are proprietary (e.g., IBM MQSeries [24] and TIBCO Rendezvous [32]), whereas other systems are deprecated and no longer supported (e.g., Gryphon [33]). A more recent work is a review paper conducted by Jiang et al. [34]. They review the current research in MOMs, present an overview of the architecture of a MOM system, and discuss some of the features of a MOM system. Jiang et al. work includes five open-source systems (e.g., Kafka and RabbitMQ). However, the paper does not explain the methodology for selecting these systems. Furthermore, the paper only provides a shallow comparison between these systems. Both of the previous papers present a high-level discussion of some features (e.g., reliability guarantees, delivery semantics, fault tolerance, and transaction support). Nevertheless, they do not discuss multi-tenancy, resource management, flow control, or active messaging. Comparing our findings with those of Eugster et al., we find that the MOM systems we study are more complex, provide richer sets of features, and offer a significantly larger number of configuration options.

Several recent efforts survey applications of MOM systems. Sheltami et al. [35] conducted a survey of pub/sub middleware solutions for wireless sensor networks. Their paper surveys existing pub/sub solutions that specifically target wireless sensor applications, such as Mires [36], TinyCOPS [37], and TinyMQ [38]. Our work differs from their survey as we focus on general-purpose MOM systems that are more complex and provide more features compared to the systems studied in [35].

Liu et al. [39] and Razzaque et al. [40] discuss the role of MOM systems in IoT applications. Liu et al. [39] first discuss the general architecture and the main features of a MOM system, and then propose a high-level architecture of an IoT system that employs a MOM system to exchange messages between devices. Razzaque et al. [40] present a detailed discussion of IoT characteristics and requirements and then evaluate existing middleware systems against these requirements. Their paper studies different types of middleware systems, including service-oriented, virtual-machine-based, event-based, and agent-based middleware systems. Their survey provides a high-level summary of each system and does not discuss many of the features we study. Furthermore, none of the systems we study was included in the aforementioned surveys [35], [40].

### B. Benchmarking Studies

The community has developed a number of benchmarks for MOM systems, including SPECjms2007 [41], jms2009-PS [42], and OpenMessaging benchmark [43]. SPECjms2007 and jms2009-PS are geared toward benchmarking MOM systems that are based on the Java Messaging Service (JMS) protocol. OpenMessaging benchmark targets evaluating a set of MOM systems on cloud platforms. Ahuja et al. [44] compares these benchmarks in terms of configurability, scalability, supported workload, and portability.

A few recent efforts focus on quantitatively evaluating the performance of open-source MOM systems. Jain et al. [45] compare the performance of Apache Pulsar and NATS using OpenMessaging benchmark [43], whereas John et al. [46] compare the performance of Apache Kafka and RabbitMQ using Flotilla [47]. These two papers perform only empirical evaluation of these systems in terms of throughput and latency and do not provide an in-depth qualitative comparison.

Our work complements these benchmarking efforts as we focus on qualitatively comparing the characteristics of popular open-source MOM systems.

## III. METHODOLOGY

We conduct an in-depth study of ten diverse, widely popular, and general-purpose open-source MOM systems. The systems we study are highlighted in Table I. They adopt different topologies, support a wide range of features, and are mature. We exclude proprietary MOM systems and cloud services such as Google Cloud Pub/Sub [27], Amazon Simple Queue service [28], IBM MQ [24], and Microsoft Azure Queue Storage [29].

We select the ten systems we study using the following methodology. First, we use the GitHub search API [81] to search for open-source MOM systems. We use keywords such as "messaging," "message queuing," "broker," "publish subscribe," and "message bus." Appendix A includes the full list of keywords we used. Using this approach, we found 57,768 GitHub projects.

Second, we choose all the projects that had 300 stars or more. Star counts are often used as an indicator of a project's popularity [82]. This reduced the number of projects to 375. Third, we manually inspect the projects and exclude projects that do not build MOM systems (e.g., applications or project specific system, i.e., not general-purpose). This reduced the number of projects to 72.

Fourth, we exclude 21 inactive projects. We consider a project to be inactive if the documentation states that the project is inactive or if there were no commits to the project in the prior year [83]. This reduced the number of projects to 50. Fifth, we exclude ten projects because they have limited documentation or the documentation is not in English. This left us with 41 projects.

Sixth, we inspect the remaining 41 projects and exclude 16 projects because they do not offer a readily-deployable MOM system but provide libraries for building custom messaging applications (e.g., ZeroMQ [84] and Nanomsg [85]). We exclude 6 projects that implement a specialized MOM system for job scheduling. Table I lists the remaining MOM projects. Table I shows two rows for Redis because it offers two MOM systems, Redis Streams and Redis PubSub.

Seventh, we examine the systems listed in Table I to identify how they disseminate messages. We identified four common dissemination topologies (Section V): peer-to-peer, single broker, mesh, and flexible topologies. For our in-depth analysis, we select the two projects that represent each topology with the largest number of stars. Although Redis Stream and Redis PubSub share a code base, we study them separately because they have different characteristics. ActiveMQ has two distributions: Classic and Artemis. ActiveMQ Classic is the older and most widely used distribution; nevertheless, the community is working on phasing out Classic and replacing it with Artemis. As a result, we study the two ActiveMQ distributions. NATS offers a "Core" deployment option that offers core features and a "JetStream" extension that add additional features atop the "Core" implementation. We study both NATS deployment options.

The selected systems are widely popular and used in production by major services. For instance, Ejabberd is used by WhatsApp [86] and Ubisoft [87], Kafka by LinkedIn [88] and Lyft [89], RabbitMQ by SoundCloud [90], and Pulsar by Yahoo [91]. Although we study the open-source versions of these systems, most of the systems we examine have a proprietary enterprise (or supported) version.

For each of the ten selected systems, we study their design documents, administrator and user manuals, tutorials, API specifications, developer blogs, and available publications. For some systems, the documentation was not complete, and we had to search the users and developers' forums, ask the system developers, and inspect the source code to determine detailed information about some characteristics.

Our process involved multiple iterations to learn about the systems. In early iterations, we identify new characteristics or

TABLE I

The short list of systems listed in descending order of the number of GitHub stars. E indicates that there is an enterprise version or support for the project. The Release Date is the date of the first open-source release of the project (sometimes a project starts as a propriety system and is later released as open-source). The shaded rows are the ten systems we study in depth.

| Project Name | License | Proprietary | Language | Version | Topologies | | | | Github Data | |
| --- | --- | --- | --- | --- | --- | --- | --- | --- | --- | --- |
| | | | | | P2P | Single | Mesh | Flexible | Stars | Release Date |
| Redis Streams [48] | BSD | E [49] | C | 6.2.5 | | ✓ | ✓ | | 51604 | May 2018 |
| Redis PubSub [50] | BSD | E [49] | C | 6.2.5 | | ✓ | ✓ | | 51604 | May 2010 |
| NSQ [51] | MIT | | Go | 1.2.1 | ✓ | | | | 20449 | Oct 2012 |
| Kafka [20] | Apache | E [52] | Java | 3.0 | | ✓ | | | 20237 | Jan 2012 |
| RocketMQ [53] | Apache | E [54] | Java | 4.8.0 | | ✓ | | | 15735 | Jan 2017 |
| NATS[1] [55] | Apache | E [56] | Go | 2.6.1 | | ✓ | ✓ | | 10033 | Jun 2014[2] |
| Pulsar [21] | Apache | E [57] | Java | 2.8.0 | | ✓ | ✓[3] | | 9795 | Aug 2016 |
| RabbitMQ [22] | MPL | E [58] | Erlang | 3.8.17 | | ✓ | ✓[4] | ✓[5] | 8884 | Aug 2007 |
| EMQX [59] | Apache | E [60] | Erlang | 4.3 | | ✓ | ✓ | | 8723 | Dec 2012 |
| Mosquitto [61] | EPL/EDL | E [62] | C | 2.0.12 | | ✓ | | | 5651 | Dec 2009 |
| Ejabberd (eCS) [63] | GPL | E [64] | Erlang | 21.07 | ✓ | ✓ | ✓ | | 5033 | Nov 2003 |
| FAYE [65] | Apache | | JavaScript | 1.4.0 | | ✓ | | | 4355 | Jun 2009 |
| Emitter [66] | AGPL | E [67] | Go | 2.8 | | ✓ | ✓ | | 3082 | Jun 2019 |
| NCHAN [68] | MIT | | C | 1.2.12 | | ✓ | | | 2777 | Nov 2009 |
| VerneMQ [69] | Apache | E [70] | Erlang | 1.12.3 | | ✓ | ✓ | | 2594 | May 2015 |
| ActiveMQ Classic [71] | Apache | E [72] | Java | 5.16.3 | ✓ | ✓ | ✓ | ✓[6] | 1953 | Jan 2007 |
| Aedes [73] | MIT | E [74] | JavaScript | 0.46.1 | | ✓ | | | 1198 | Mar 2015 |
| ActiveMQ Artemis [23] | Apache | E [72] | Java | 2.17.0 | ✓ | ✓ | ✓ | ✓[7] | 744 | May 2015 |
| HiveMQ CE [75] | Apache | E [76] | Java | 2021.2 | | ✓ | | | 672 | Apr 2019 |

(1) NATS refers to both NATS Core and NATS JetStream.
(2) NATS JetStream was first released on March 2021.
(3) The mesh topology is created when using geo-replication.
(4) A mesh topology is created when using non-exclusive queues.
(5) Supported through the Federation [77] or the Shovel [78] plugins.
(6) Supported through the concept of Virtual Destinations [79].
(7) Supported by coupling Federation Addresses with Divert Bindings [80].

refine our definition of the characteristics we use to compare systems. In the end, we identified 42 distinct characteristics with a total of 134 different options.

### A. Limitations

As with any in-depth study, the list of properties and system designs may not represent all available systems. Here, we list four potential sources of bias and describe our best efforts to address them.

- *Representativeness of the studied systems*: Although we study ten diverse systems (highlighted in Table I), our results may not generalize to systems we did not study. To ensure the representativeness, we followed a rigorous methodology to short listing 19 open-source, popular, diverse, and general-purpose MOM systems.
- *Limiting the study to systems with 300+ stars*: Although we only consider projects with 300 stars or more, this choice does not impact the final set of selected systems. We choose the most popular system for each dissemination topology; consequently, considering systems with lower than 300 stars would not have changed the set of projects we study.
- *Documentation accuracy and completeness*: Our team predominantly studied the publicly available sources of documentation of the selected systems. We did not study the complete source code nor deploy all the selected systems. Consequently, the accuracy of our findings inherently depends on the accuracy and completeness of the available documentation.
- *Observer error*: We study 42 characteristics in ten systems. Some of the characteristics are not explicitly documented in the available documentation and require a careful reading to determine whether a system supports a specific characteristic. To reduce the chance of observer errors, two team members study each system independently following the same methodology, then compare their findings. If a disagreement arises, the entire team convenes in a group meeting to discuss the disagreement before reaching a verdict.

## IV. Message-Oriented Middleware Communication Model

One challenge in conducting an in-depth study of MOM systems is that the systems we study do not agree on the core features a MOM system should offer Furthermore, the community does not use standard terminology for describing the various aspects of MOM systems. Worse yet, sometimes the same term has different meanings in different systems. For further details, please refer to Note 1.

## A. MOM Communication Model

We now present an abstract communication model that we use to capture the basic communication functionality offered by the majority of the systems we study. Given that there is no standard terminology for describing MOM systems, we also define the terminology we use in this work.

The communication model offers a communication abstraction for producers (i.e., senders or publishers) to send messages to one or more consumers (i.e., receivers or subscribers). A message carries application-specific data. MOM systems usually rely on a broker service to exchange messages. Producers send their messages to a broker service, which in turn forwards them to interested consumers.

The MOM systems we studied rely on a broker service to exchange messages, with the exception of NSQ, which is a peer-to-peer system. Communication centers on topics (also referred to as *subjects*, *addresses*, and *channels*.), which have unique IDs that can be human-readable. Producers send messages to the broker service for a certain topic and consumers subscribe to that topic to receive messages. A typical application that uses a MOM system involves multiple producers and consumers that communicate through application-specific topics. Applications support multiple users, where a user may run multiple producers and consumers to produce and consume messages to and from different topics.

The broker hosts one or more logical *exchanges*. An exchange is either a multicast exchange, in which a message is delivered to all of the consumers that subscribed to that exchange, or a unicast exchange, in which a message is only delivered to one of the consumers subscribed to that exchange. Multiple consumers may subscribe to a unicast exchange. However, each message sent to a unicast exchange is delivered to only one of the subscribed consumers. MOM systems use a range of policies to select a consumer among those that are subscribed to a unicast exchange. Examples of these policies include random, round-robin, priority-based, and First-Come-First-Served (FCFS) policies. In Section X-C, we detail the policies offered by the systems we study. Figure 1 shows an example application that has multiple producers and consumers communicating through a single topic with the name "UW-news." The broker has a single multicast exchange.

MOM systems (e.g., RabbitMQ, ActiveMQ Classic, and ActiveMQ Artemis) allow the use of multiple exchanges to form what we call a *logical dissemination graph* or *logical topology*. Figure 2 shows an example in which a single broker hosts a logical topology that uses two multicast exchanges and three unicast exchanges.

The logical topology can be deployed on one or more brokers. We refer to how a logical topology is placed on the brokers as the *physical layout*. Figure 3 shows the same logical topology in Figure 2 when it is laid out on two broker servers across two data centers.

## B. MOM Communication Characteristics

The characteristics of the MOM communication model are drastically different from other popular communication middleware, such as raw TCP/IP sockets and RPC (e.g.,

> **Note 1: Terminology in Real Systems**
> We note that different systems use different names to refer to the concepts presented in our model. In some cases, the same term has different meanings in two systems. For example, a multicast exchange is called a "topic" in NSQ, Kafka, RocketMQ, Pulsar, EMQX, Mosquitto, VerneMQ, ActiveMQ Classic, Aedes, and HiveMQ; "topic exchange" in RabbitMQ; "subject" in NATS; "stream" in Redis Streams; and "channel" in Redis PubSub, FAYE, Ejabberd, Emitter, and NCHAN. Interestingly, a "channel" in NSQ refers to a unicast exchange instead of a multicast exchange. A unicast exchange is called a "queue" in RabbitMQ and ActiveMQ Classic. ActiveMQ Artemis exchanges are called "addresses", with each "address" has an attribute that specifies whether it is a multicast or unicast exchange.

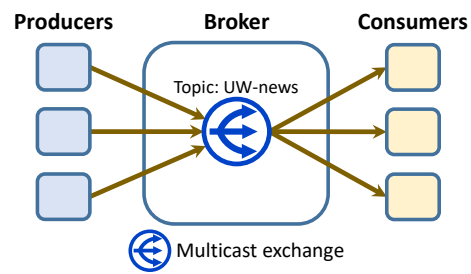

Fig. 1. A single broker with a multicast exchange. Producers send massages to the "UW-news" topic. The broker multicasts the messages to all consumers.

Linux RPC [92], Google gRPC [93], and Apache Thrift [94]). These characteristics make it attractive for modern cloud applications:

- *Space decoupling*: Producers and consumers do not need to be aware of each other. A producer does not have the addresses of the other producers or consumers, nor know how many of them are there. Similarly, consumers do not know the address or the number of other consumers and producers.

- *Time decoupling*: MOM systems often facilitate the time decoupling of producers and consumers, which means that producers and consumers do not need to interact synchronously to communicate. Producers can send messages to a MOM system while a consumer is offline. The consumer will receive the messages when it joins the system, even if the producer is offline.

- *Flexibility*: MOM systems can be configured to provide a range of reliability, resource management, and delivery semantics. They can be deployed to match the underlying infrastructure to improve performance and scalability, such as placing brokers on central nodes or configuring a physical layout for geographically distributed setups to reduce WAN communication.

- *Separation of concerns and extensibility*: Producers and consumers communicate using a simple messaging API without being concerned with the way this API is implemented. Consequently, it is easy to extend the system design, change its implementation to add new features, or improve its performance without changing the imple-

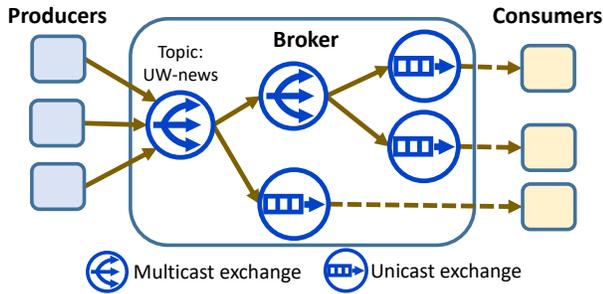

Fig. 2. A single broker hosting a logical topology with multiple exchanges. The arrows show the flow of messages.

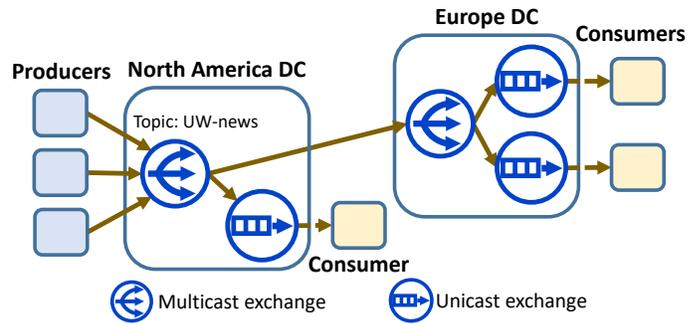

Fig. 3. Flexible topology. A logical topology deployed on two physical brokers located in two data centers.

**Note 2: How our Communication Model is Mapped to Different Systems**

To demonstrate how our model can be translated to deployments in the systems we study, consider a load balancing use case in which a message should be processed by one of the database servers and one of the web servers (Figure 4). Producers send messages to the topic "TX" that is processed by a multicast exchange. The multicast exchange sends a copy of the message to the unicast exchanges of the database and web servers. Each unicast exchange forwards the request to one of the servers subscribed to that exchange. The systems we study use different approaches to implement this use case. We detail how this use case can be implemented in two systems: RabbitMQ and Redis Streams.

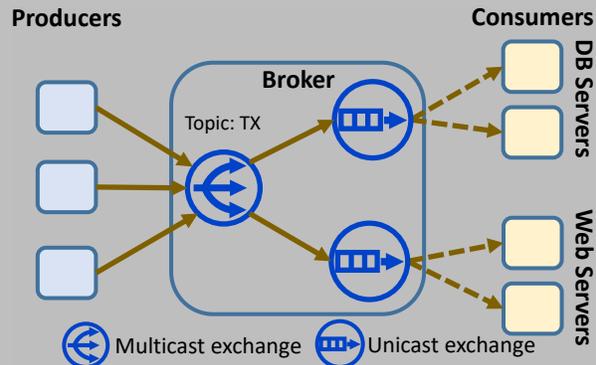

Fig. 4. A use case of load balancing a single multicast exchange.

One can directly implement the described use case in RabbitMQ. A multicast exchange is called a *topic exchange*, while a unicast exchange is called a *queue*. Multiple consumers can subscribe to a queue. Messages sent to a queue are forwarded using one of the RabbitMQ policies listed in Table XIV. NSQ can implement this use case in a similar way.

Redis Streams does not have explicit unicast exchanges. Each message produced to a multicast exchange (also known as a *stream*) is available to all consumers of that stream to request the message. The functionality of a unicast exchange can be achieved through a feature called *consumer groups*. Redis Streams supports grouping multiple consumers into a consumer group. A consumer group can have a single subscription to a multicast exchange, and messages sent to the group subscription are load-balanced among the group consumers. To implement the use case in Figure 4 using Redis Streams, one would define two consumer groups, one for database servers and one for web servers, and then subscribe each group to the multicast exchange serving the topic "TX." Many systems implement the unicast exchange using this approach including Pulsar and NATS. Consumer groups are called *shared subscriptions* in Pulsar. We note that shared subscriptions became a standard feature in version 5.0 of the MQTT messaging protocol.

mentation of the communicating parties.

## V. MOM TOPOLOGIES

MOM systems use a dissemination topology to transfer messages to consumers. We found that the systems we study support one or more of the following four dissemination typologies (Table II): single broker, mesh, peer-to-peer, and flexible topology.

### A. Single Broker Topology

A single broker is the simplest and most popular dissemination topology, in which a single node runs the MOM service

(Figure 1). The broker hosts all topics, receives all messages, and serves all consumers. This topology is supported by all the systems we study with the exception of NSQ, which is a brokerless peer-to-peer system (Table II).

Although using a single broker server simplifies the design and implementation of the MOM system, it introduces a single point of failure. This is why systems adopting this single-broker topology may replicate the state of the broker to backup brokers. For instance, Figure 5 shows how messages are replicated in Kafka, where topic partitions are replicated on multiple brokers, with one of the brokers acting as the primary replica for a specific topic partition. Producers send

TABLE II
DISSEMINATION TOPOLOGIES.

| System | | Redis Streams | Redis PubSub | NSQ | Kafka | NATS | Pulsar | RabbitMQ | Ejabberd (eCS) | ActiveMQ Classic | ActiveMQ Artemis |
|---|---|---|---|---|---|---|---|---|---|---|---|
| Topology | Single | ✓ | ✓ | | ✓ | ✓ | ✓ | ✓ | ✓ | ✓ | ✓ |
| | Mesh | ✓ | ✓ | | | ✓ | ✓[1] | ✓[2] | ✓ | ✓ | ✓ |
| | Flexible | | | | | | | ✓[3] | | ✓[4] | ✓[5] |
| | Peer-to-Peer | | | ✓ | | | | | ✓ | ✓ | ✓ |

(1) The mesh topology is created when using geo-replication.
(2) A mesh topology is created when using non-exclusive queues.
(3) Supported through the Federation [77] or the Shovel [78] plugins.
(4) Supported through the concept of Virtual Destinations [79].
(5) Supported by coupling Federation Addresses with Divert Bindings [80].

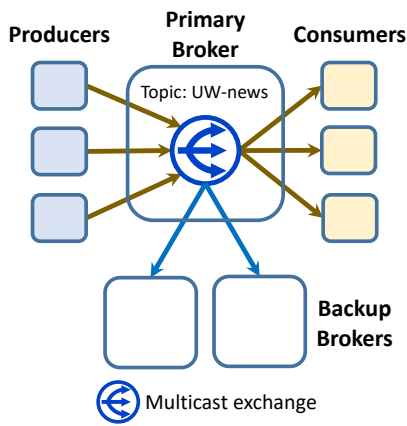

Fig. 5. Single broker topology example. The figure shows how Kafka replicates the partitions of a topic on multiple brokers.

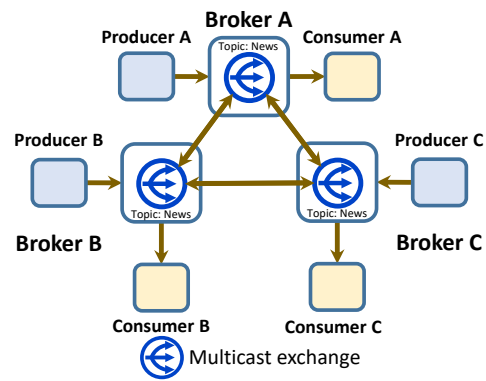

Fig. 6. Complete mesh topology. A topic is served by three multicast exchanges, each deployed on a separate broker. The brokers are connected in a complete mesh topology.

messages to the primary broker of a given topic partition (also known as the *leader*). The primary broker can be configured to replicate every message to backup brokers before making it available to consumers. The backup brokers do not serve consumers. One of the backup brokers takes over the primary role when the primary broker fails. Note that a backup broker may act as a primary broker for other topic partitions.

The main disadvantage of this design is that its scalability and performance are limited to a single server's capacity. Furthermore, this topology is ill-suited for multiple data center deployments.

### B. Complete Mesh Topology

In a complete mesh topology, every broker is connected to all other brokers. The same topic is served by all the brokers. The main goal of the mesh topology is to increase system scalability and to improve system throughput. Figure 6 shows an example of a complete mesh topology with three brokers each hosting a multicast exchange that serves the same topic, called "News." As a result, a consumer can subscribe to any of the brokers. When a broker receives a new message from a producer, it forwards the message to all other brokers, which in turn forward this message to their local consumers, who are subscribed to that topic. RabbitMQ further optimizes this approach by only forwarding messages to brokers that have at least one consumer subscribed to that topic. This topology is supported by eight of the systems we study (Table II) including: Redis Streams, Redis PubSub, NATS, Pulsar, RabbitMQ, Ejabberd, ActiveMQ Classic, and ActiveMQ Artemis.

### C. Flexible Topology

RabbitMQ, ActiveMQ Classic, and ActiveMQ Artemis support building a logical message dissemination topology and flexibly deploying it on multiple brokers (e.g., Figure 3). Each broker can host a subgraph of the logical topology. This is the most flexible design, and it can be configured to support single-broker and complete mesh topologies.

Furthermore, this flexibility allows matching the dissemination topology with the underlying network within a data center or across data centers. This also helps to support a wide range of application-specific topologies and allows systems to scale to support a larger number of producers and consumers. This flexibility introduces additional steps for deploying these

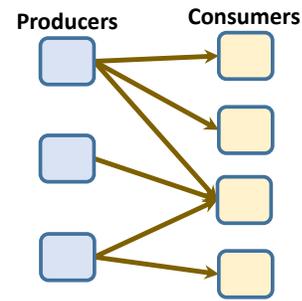

Fig. 7. Peer-to-peer topology. Consumers connect directly to producers without an intermediary broker node.

---

> **Note 3: Flexible Topology Implementation Overview**
>
> In RabbitMQ, flexibility is achieved through the *Federation* [77] or the *Shovel* [78] plugins. RabbitMQ has *topic exchanges*, which act as multicast exchanges, and *queues*, which act as unicast exchanges. The Federation plugin can forward messages between topic exchanges and queues, even if they reside on different brokers. Furthermore, multiple topic exchanges and queues can be used to create a variety of dissemination topologies. The Shovel plugin is similar to the Federation plugin, except it works on a lower level to consume messages from queues to topic exchanges.
>
> In ActiveMQ Classic, a flexible topology can be created using *virtual destinations* [80]. A topic can be configured as a virtual destination to enable consuming messages from other topics. This added flexibility allows the creation of flexible dissemination topologies.
>
> In ActiveMQ Artemis, flexibility is achieved through the use of *federated addresses*. An address is a topic in our terminology. An address can be configured as a multicast or unicast address. Through the *divert bindings* feature [80], a multicast or unicast address subscribes to one or more multicast source addresses. Multiple addresses can be used to create a flexible dissemination topology.

systems because it requires configuring and tuning the system for each new deployment.

### D. Brokerless Peer-to-Peer Topology

This topology does not use dedicated nodes to run the broker service. Producers create topics on their local machines and consumers directly subscribe to these topics, without using an intermediary broker service (Figure 7). These systems use a discovery service to help consumers find the producers that hosts a certain topic. If more than one producer is producing to the same topic, a consumer needs to independently subscribe to each of these producers. With this approach, a consumer will receive all the messages sent to a given topic, but the order in which messages are received may differ between consumers of the same topic.

The main advantages of this topology are its high availability, low communication latency, and the elimination of dedicated broker nodes. It has high scalability for applications that are dominated by one-to-one communication. Consequently, this topology is often used in real-time communication applications such as instant messaging [86], gaming [87], VoIP calls [95], and video calls [96].

Unfortunately, this topology has serious limitations. First, this approach cannot scale to support hundreds of consumers that subscribe to the same producers of a topic. Second, this approach eliminates a number of important MOM communication characteristics (Section IV-B) including: time decoupling because producers and consumers communicate directly, space decoupling because producers and consumers know each other, and flexibility. Finally, this design limits the number of features that can be supported, such as replication and message ordering per multicast exchange.

NSQ is a pure peer-to-peer system. ActiveMQ Classic and ActiveMQ Artemis offer libraries that allow the integration of the MOM service into producers and consumers to create a peer-to-peer system. Additionally, Ejabberd allows producers and consumers to establish direct peer-to-peer connections for lower latency.

### E. Summary

In this Section, we present the four message dissemination topologies that MOM systems use to transfer messages from producers to consumers. Among these topologies, the single broker topology is the most supported and simplest topology, but it introduces a single point of failure. Therefore, replicating the broker to replicas or using complete mesh topology instead is recommended for scalability and reliability. P2P-based MOM systems lack the support of most of the characteristics of the MOM communication model (Section IV-B). Interestingly, some systems support building custom message dissemination topologies that match the underlying network infrastructure and support complex access patterns including combinations of single broker and complete mesh topologies. Finally, Table II shows that most systems support two or more topologies with ActiveMQ Classic and Artemis supporting all possible dissemination topologies.

## VI. RELIABILITY

An important consideration when designing, implementing, and deploying MOM systems is the service and data durability guarantees that they provide during failures. The MOM systems that we study incorporate techniques to tolerate broker and consumer failures, including data durability (Section VI-A), replication (Section VI-B), reassigning messages of failed consumers (Section VI-C), and consumer subscription recovery (Section VI-D).

### A. Data Durability

The core mechanism to achieve data durability is to store messages on a stable storage, through a local file system, a database, or a key-value storage system. Table III lists the systems that provide data durability. We note that all systems support storing data to durable storage except Redis PubSub and NATS Core. These two systems are designed for best effort

TABLE III
Message persistence, granularity of persistence, and persistence frequency for each of the systems we study. J refers to a feature exclusively supported by NATS JetStream. Although Pulsar and Ejabberd can use different storage engines by using plugins, the default option is BookKeeper [97] for Pulsar and the Mnesia [98] database for Ejabberd. In this table, we report findings based on Pulsar and Ejabberd's default implementation.

| System | | Redis Streams | Redis PubSub | NSQ | Kafka | NATS | Pulsar | RabbitMQ | Ejabberd (eCS) | ActiveMQ Classic | ActiveMQ Artemis |
|---|---|---|---|---|---|---|---|---|---|---|---|
| Data Durability | Persistent messages | ✓ | | ✓[1] | ✓ | ✓ J | ✓ | ✓ | ✓ | ✓ | ✓ |
| | Non-persistent messages | ✓ | ✓ | ✓ | | ✓ | ✓ | ✓ | ✓ | ✓ | ✓ |
| Persistence Granularity | System-wide | ✓ | | ✓ | ✓ | ✓ J | ✓ | ✓ | ✓ | ✓ | ✓ |
| | Per Topic | | | | | | ✓ | ✓ | ✓[2] | ✓ | ✓ |
| | Per Message | | | | | | | ✓ | | ✓ | ✓ |
| Persistence Frequency | Immediate | ✓ | | | ✓ | | | | | ✓ | ✓ |
| | Periodic | ✓ | | | ✓ | ✓ J | ✓ | ✓ | ✓ | ✓ | ✓ |
| | Batching | | | | ✓ | | ✓ | ✓ | ✓ | | ✓ |
| | Operating System Persistence | ✓ | | ✓ | ✓ | ✓ J | ✓ | | | ✓ | ✓ |

(1) By default, NSQ is an in-memory messaging system that persists messages to disk if the size of messages in memory exceeds a threshold. Setting this threshold to zero forces persisting messages to disk immediately.
(2) Only when using Ejabberd PubSub module.

delivery, where a message is removed once it is dispatched to the subscribed consumers.

Kafka is the only system that stores every message to local storage and does not offer the ability to disable durable data storage. The rest of the systems include configuration options to disable storing data durably for all or a subset of messages.

**The frequency of making messages durable.** Storing a message to a stable storage imposes an overhead. To avoid performing I/O operations on the critical path, many of the systems allow users to tune the frequency with which messages are synced to the durable storage. Among the systems we study, we found four frequency configurations:

- *Immediate:* This configuration stores messages at the stable storage by calling `fsync()` after every publish request. An acknowledgement is sent to the producer only after `fsync()` returns. This configuration offers the highest durability guarantee; however, it imposes the highest overhead because `fsync()` can incur significant overhead and it is called on the critical path.
- *Periodic:* Using this configuration, an `fsync()` is issued every preconfigured period of time. The shorter the period of time, the lower the chance of losing data and the higher the performance overhead. Using very long periods is not advised because it lowers the system reliability and introduces I/O bursts.
- *Batching:* Messages are made persistent when the system accumulates a preconfigured number or total bytes of messages. Similar to the periodic configuration, the smaller the batch size, the higher the system reliability and the higher the imposed overhead.
- *OS Persistence:* This configuration leaves the decision to write messages to disk to the local file system. The MOM system writes the data to the file system without calling `fsync()` and the file system eventually writes the data to disk.

With the exception of NSQ, periodic durability is supported by all systems that support durable message storage. OS-based persistence is supported by all systems with the exception of RabbitMQ and Ejabberd, which use the internal Mnesia database for durability (Table III). Redis Streams, Kafka, ActiveMQ Classic, and ActiveMQ Artemis are the only systems that support immediate durability. Finally, a number of systems including Kafka, Pulsar, RabbitMQ, Ejabberd, and ActiveMQ Artemis offer a hybrid approach that combines periodic and batch-based policies. These systems persist a message if a certain period passes or the number of accumulated messages exceeds a configurable threshold.

**Granularity of the Durability Configuration.** The durability configuration can be set as a system-wide configuration, per topic, or per message. All systems that support persistent messages can be configured to store every message produced (Table III). Kafka is the only system that does not offer an option to disable storing messages on persistent storage; it only provides configuration options to tune the mechanism. Instead of durably storing all messages, some of the systems offer two finer granularity levels: topic and message.

Topic durability in Pulsar, RabbitMQ, Ejabberd PubSub, ActiveMQ Classic, and ActiveMQ Artemis allows for the configuration of a topic as either durable or transient. This granularity supports the durable storage of messages of a certain topic.

Message-level durability is the finest granularity offered for data durability. RabbitMQ, ActiveMQ Classic, and ActiveMQ Artemis allow one to specify the durability configuration per message. The producer API includes a parameter that indicates whether a given message should be durably stored.

In systems that offer topic-level and message-level durability, if a transient message is sent to a durable topic, or if a message with a durability parameter set to true is produced to a nondurable topic, the message will not be durably stored.

TABLE IV
REPLICATION ALTERNATIVES. PULSAR AND EJABBERD ALLOW PLUGGING IN STORAGE ENGINES, THE RESULTS IN THE TABLE ARE BASED
ON THE DEFAULT OPTIONS OF BOOKKEEPER IN PULSAR, AND THE MNESIA DATABASE IN EJABBERD.

| System | | Redis Streams | Redis PubSub | NSQ | Kafka | NATS JetStream | Pulsar | RabbitMQ | Ejabberd (eCS) | ActiveMQ Classic | ActiveMQ Artemis |
|---|---|---|---|---|---|---|---|---|---|---|---|
| Replication | Supported | ✓ | | | ✓ | ✓ | ✓[1] | ✓[2] | | | ✓ |
| | Provided by storage engine | | | | | | ✓ | | ✓ | ✓[3] | ✓[3] |
| | Rack-Aware | | | | ✓ | | ✓ | | | | |

(1) Supported through the geo-replication feature.
(2) Exclusive to the special replicated queue which is called Quorum Queue.
(3) Provided through a shared file system.

> **Note 4: Replication implementation in RabbitMQ, NATS JetStream, and Kafka**
>
> RabbitMQ and NATS JetStream build replication techniques based on the Raft linearizable consensus protocol [99]. NATS JetStream creates multiple Raft instances. One global Raft instance is used to keep track of brokers in the system, one Raft group per topic is used to track brokers serving that topic, and one Raft group per topic is used for the consumers of a given topic.
>
> Kafka builds its own replication protocol [20] and it partitions topics onto different brokers. For each partition, Kafka assigns one primary broker that serves all client requests. When the primary broker receives a new message, it replicates the message on backup brokers before serving it to consumers. Kafka can be configured to place replicas of a partition on different racks for higher reliability. Although immediate persistence to disk by calling `fsync()` after every publish request can be combined with replication on backup nodes, Kafka's documentation discourages this configuration due to performance implications.

## B. Replication

Table IV shows that all broker-based systems except Redis PubSub and NATS Core support message replication for reliability, or scalability to support larger deployments. Redis Streams, Kafka, NATS JetStream, Pulsar, RabbitMQ, Ejabberd, and ActiveMQ Artemis natively support message replication. Furthermore, Pulsar, Ejabberd, ActiveMQ Classic, and ActiveMQ Artemis can be configured to use external replicated storage. For higher reliability, Kafka and Pulsar can be configured to follow a rack-aware replica placement. Rack-aware placement aims to place replicas on different racks to tolerate rack failures.

## C. Consumer Fault Tolerance

In MOM systems, consumers can be configured to send an acknowledgment when they finish processing the received message. This is the case for all the systems that we study with the exception of Redis PubSub and NATS Core. If a consumer receives a message from a MOM system but it fails or its TCP connection drops before acknowledging it, then the MOM system assumes that the message has not been processed and it reassigns the message to another consumer.

We note that this fault tolerance technique is only provided for unicast exchanges, because each message sent to a unicast exchange is forwarded to one of the consumers who are subscribed to that exchange. If a consumer fails, the in-flight message(s) should be reassigned to another consumer. If no other consumer exists, the unicast exchange keeps the message(s) until a consumer subscribes to the exchange. This is not the case for multicast exchanges, which forward every message to all consumers subscribed to that multicast exchange. Table V lists the reassignment policies supported by the systems we study. These policies are as follows.

- **Reassigned by the broker**: In this policy, the broker reassigns the unacknowledged messages of a failed consumer to another consumer using one of the unicast exchange policies discussed in Section X-C. This policy is supported by NATS JetStream, Pulsar, RabbitMQ, ActiveMQ Classic, and ActiveMQ Artemis.

- **Reassigned by a consumer**: Kafka and Redis Streams allow an application to decide which consumer will claim the messages assigned to a failed consumer. In Kafka, a topic is divided into multiple partitions. A consumer claims complete partitions and processes all messages assigned to those partitions. When a consumer fails, its partitions need to be reassigned to other consumers. This reassignment can be done by the system or by the application. In the latter case, the application decides which consumer will claim the partitions of a failed consumer. In Redis Streams, consumers within the same consumer group who consume messages from a stream can claim each other's unacknowledged messages. Consequently, if a consumer within the group fails, other consumers of that group can claim unacknowledged messages of the failed consumer. A second use of this technique in Redis Streams is to mitigate slow consumers. A fast consumer can claim unacknowledged messages of a slow consumer. Unfortunately, this approach may lead to double processing of a message.

The reassignment of messages by consumers is also supported by Pulsar, RabbitMQ, ActiveMQ Classic, and

TABLE V
The supported recovery guarantees in case of consumer failure.

| System | | Redis Streams | Redis PubSub | NSQ | Kafka | NATS JetStream | Pulsar | RabbitMQ | Ejabberd (eCS) | ActiveMQ Classic | ActiveMQ Artemis |
|---|---|---|---|---|---|---|---|---|---|---|---|
| Consumer Fault Tolerance | Reassigned by the Broker | | | | | ✓ | ✓ | ✓ | | ✓ | ✓ |
| | Reassigned by a Consumer | ✓ | | | ✓ | | | | | | |
| | Requeue as a new | | | ✓ | | | | | | | ✓¹ |
| Subscription Recovery for Multicast Only | | ✓ | | | ✓ | ✓ | ✓ | ✓² | ✓³ | | ✓ |

(1) Applies only to ring queues.
(2) Using an exclusive queue or a durable subscription.
(3) Supported when enabling the "mod_offline" or "mod_mam" modules.

ActiveMQ Artemis but not for consumer fault tolerance. These systems place messages that a consumer could not process (e.g., because of a violation of the application semantics) into a "dead letter" topic. The application can use a consumer to inspect these messages.

- *Requeue as new*: Unacknowledged messages of a failed consumer are requeued as new messages without keeping track of the previous delivery attempts. The message is added to the tail of the unicast exchange and it can be forwarded to the same failed consumer if it rejoins the system. This policy is only supported by NSQ and ActiveMQ Artemis.

### D. Subscription Recovery

With the exception of Redis PubSub, NSQ, and NATS Core, all the systems we study offer a technique to recover old subscriptions if a consumer fails (Table V). Recovering a subscription includes reinstating the subscription to its previous state before the crash.

Subscription recovery is provided only for multicast exchanges because a consumer of a multicast exchange should receive a copy of every message sent to that exchange. This recovery technique is not needed for unicast exchanges. A consumer of a unicast exchange might not receive all the messages sent to the exchange when multiple consumers exist, even when there are no failures. When a consumer of a unicast exchange fails and rejoins the cluster, it joins back as a new consumer.

Subscription recovery is achieved through one of two techniques:

- *Broker-side recovery*: The broker is tasked with storing the subscription state, such as the consumer information and the last consumed message ID. The broker then reinstates the subscription when a consumer rejoins the system. This approach is supported by Kafka, NATS JetStream, Pulsar, RabbitMQ, Ejabberd PubSub, ActiveMQ Classic, and ActiveMQ Artemis. We note that two popular messaging protocols, JMS v1.0.1+ and MQTT v3.1.1+, include specifications for "Durable Subscriptions" that are reinstated when a consumer rejoins the system. We discuss messaging protocols in Section XI.

- *Client-side recovery*: The consumer stores the subscription state in persistent storage. After rejoining the system, the consumer uses the stored state to resume consuming messages from the last message ID stored on disk. This approach is supported by Redis Streams.

### E. Summary

Most of the MOM systems we study support the durable storage of messages with configurable persistence frequency. Some systems offer fine message persistence granularity, including message-level and topic-level. Other than NSQ, all systems with support for data durability support replicating messages for higher reliability or scalability, with the majority relying on built-in replication mechanisms. Although some systems support strongly consistent storage, the majority of systems offers eventually consistent storage semantics. Finally, MOM systems also incorporate techniques to tolerate consumers' failures. The majority of systems support recovering the subscription of a consumer of a multicast exchange. All systems that support the load balancing of messages of a unicast exchange across multiple consumers support the reassignment of messages of a failed consumer to another one.

## VII. MOM Service Semantics

In this section, we discuss the semantics offered by the systems we study, including message delivery semantics, message deduplication semantics, transaction semantics, and message ordering semantics.

### A. Message Delivery Semantics

We detail the delivery semantics for the transmission of messages from producer-to-broker and broker-to-consumer (or producer-to-consumer in peer-to-peer systems). The systems we study offer four types of delivery semantics. Table VI lists the semantics supported by each system.

- *Best-effort, at-most-once*: After sending a message, the sender does not wait for an acknowledgment from the receiver. If the message is lost due to failure, then no retransmission occurs. We call this technique *best-effort*,

TABLE VI
Message delivery semantics. J refers to a feature exclusively supported by NATS JetStream.

| System | | Redis Streams | Redis PubSub | NSQ | Kafka | NATS | Pulsar | RabbitMQ | Ejabberd (eCS) | ActiveMQ Classic | ActiveMQ Artemis |
|---|---|---|---|---|---|---|---|---|---|---|---|
| Producer Delivery Semantics | Best-effort at-most-once | | | | | ✓[1] | | ✓ | ✓ | ✓ | ✓ |
| | Reliable at-most-once | ✓ | ✓ | | ✓ | ✓J | ✓[3] | ✓ | ✓ | ✓ | ✓ |
| | At-least-once | | | | ✓ | ✓J[1] | ✓ | | ✓ | ✓ | ✓ |
| | Effectively-once | ✓ | | | ✓ | ✓J | ✓ | | | ✓ | ✓ |
| Consumer Delivery Semantics | Best-effort at-most-once | ✓ | ✓ | | ✓ | ✓[1] | ✓[4] | ✓ | ✓ | ✓ | |
| | Reliable at-most-once | ✓ | | | ✓ | ✓J[2] | | ✓ | ✓ | ✓ | |
| | At-least-once | | | ✓ | | ✓J[1] | ✓[4] | | | | ✓ |
| | Effectively-once | ✓[5] | | ✓[5] | ✓[5] | ✓J[5] | ✓[5] | ✓[5] | ✓[5] | ✓[5] | ✓[5] |

(1) In NATS JetStream with the MQTT Protocol, different delivery semantics are offered by different quality of service levels. QoS0 offers best-effort semantics and QoS1 offers at-least-once semantics.
(2) Applies only to poll-based consumers.
(3) The semantics depend on how the producer implementation handles acknowledgment time-outs.
(4) Topic durability configuration (Section VI-A) imposes specific delivery semantics. Nondurable topics dispatch messages as best-effort at-most-once and durable topics dispatch messages using at-least-once semantics.
(5) The implementation of effectively-once semantics is done by the application code. We added a check for systems that support transactions or provide enough information in the message to facilitate implementing message deduplication.

*at-most-once* because a message is processed at-most-once, but will be lost in case of failure.

- *Reliable, at-most-once*: After sending a message, the sender waits for an acknowledgment from the receiver. Brokers can be configured to wait synchronously or asynchronously for an acknowledgment.
- *At-least-once*: After sending a message, the sender waits for an acknowledgment from the receiver. If the sender times out before receiving an acknowledgment, the sender retransmits the message. A message could be delivered more than once. If the broker does not receive an acknowledgment from a consumer after retrying a number of times, then the message is dropped or added to a dead-letter exchange.
- *Effectively-once*: This is the strongest delivery guarantee. Similarly to at-least-once semantics, the sender retries unacknowledged messages and may deliver a message more than once. However, message deduplication techniques or atomic transactions are used to eliminate duplicate messages. We discuss these two techniques in the following sections.

### B. Handling Duplicate Messages

Messages could be retransmitted to achieve reliable delivery. This may lead to duplicate delivery of the same message from a producer to a broker (or a consumer in the case of a peer-to-peer system) or from a broker to a consumer. Here, we investigate how duplicate messages are handled both on the MOM system side and on the consumer side when the message is duplicated due to retransmissions. We do not study message duplication that is caused by the application design. For instance, an application may subscribe the same consumer twice to the same topic using two separate subscriptions, or it may use two wild-card filters that overlap (i.e., a message may match the two filters; we discuss message filtering in Section X-F).

**System side.** Table VII shows the systems that handle duplicate messages. It shows that Redis Streams, Kafka, NATS JetStream, Pulsar, ActiveMQ Classic, and ActiveMQ Artemis support the detection of duplicate messages on the broker. They identify a message as duplicate based on the message attributes, mainly the message ID or sequence number. The IDs and sequence numbers of messages are generated by their respective producers. The pair <producer ID, message ID/sequence number> represents a unique message identifier.

**Consumer side.** Handling duplicate messages on the consumer side is implemented by the application and is therefore outside the scope of our study. Table VII shows the systems that provide enough information to facilitate the implementation of a technique to handle duplicate messages on the consumer side. We found that all of the systems we studied, except Redis PubSub and NATS Core, embed enough information in messages to facilitate handling duplicate messages. Messages are uniquely identified using message IDs or unique sequence numbers. Consumers can track this information to detect duplicate messages. We note that Redis PubSub and NATS Core are in-memory best-effort systems, where retransmission is not possible and hence there is no need to support deduplication.

### C. Transactions

Some MOM systems offer transactional support for applications. Table VIII shows the systems that support transactions, the transaction properties, the supported operations, and the scope of the transaction. All systems except NSQ, NATS, and Ejabberd support transactions.

**Transaction properties.** We study the ACID properties [100] of the supported transactions. We note that unlike other properties we study, systems are not flexible and they provide a

TABLE VII
Support for message deduplication. The table reports the systems that handle duplicate messages on the system side. For the consumer side, the table reports the systems that provide information to facilitate handling duplicate messages by the application.

| System | | Redis Streams | Redis PubSub | NSQ | Kafka | NATS JetStream | Pulsar | RabbitMQ | Ejabberd (eCS) | ActiveMQ Classic | ActiveMQ Artemis |
|---|---|---|---|---|---|---|---|---|---|---|---|
| Handling Duplicate Messages | System side | $\checkmark^1$ | | | $\checkmark$ | $\checkmark$ | $\checkmark$ | | | $\checkmark^2$ | $\checkmark$ |
| | Consumer side | $\checkmark$ | | $\checkmark$ | $\checkmark$ | $\checkmark$ | $\checkmark$ | $\checkmark$ | $\checkmark$ | $\checkmark$ | $\checkmark$ |

(1) If the producer creates an ID for a message, then the broker rejects messages with an ID smaller or equal to the last seen ID.
(2) By using the "Auditing" feature, the broker maintains a sliding window of messages to detect duplicates.

specific transaction support. For instance, a system can support either atomic or non-atomic transactions, but not both.

- *Atomicity*: Pulsar, ActiveMQ Classic, and ActiveMQ Artemis offer atomic transactions such that the operations carried out within a transaction are either executed entirely or not at all. Redis Streams, Redis PubSub, Kafka, and RabbitMQ offer non-atomic transactions. In these non-atomic transactions, partial results could be visible due to either system failures in RabbitMQ, lack of support for rollback of produced messages in Kafka, or failures of execution of some of the operations within a transaction in Redis Streams and Redis PubSub.
- *Consistency*: MOM systems do not check data consistency. MOM systems aim to be generic and capable of supporting a wide range of applications. Consequently, MOM systems do not check the schema of the message payload. Data consistency semantics are application-specific and are left to the application to handle.
- *Isolation*: MOM systems' transactions offer classic transaction isolation, in which concurrent transactions are serializable. In other words, the intermediate results of a running transaction are not visible to other transactions until the transaction is committed. Among systems that support transactions, Kafka is the only system that does not support isolation. In Kafka, produced messages are immediately stored in the log, messages from concurrent producers are mixed in the log, and the system does not support rolling back produced messages. Consequently, consumers outside a transaction may read messages of that transaction before it commits, or even after it aborts.
- *Durability*: Transaction durability ensures that the changes of a committed transaction remain in the system even in the case of failure. We note that systems that support transactions typically do not have a separate configuration option for transaction durability. Instead, transaction durability follows the durability configuration discussed in Section VI-A. For instance, if a transaction that uses durable messages is committed on a durable topic, then the transaction is durable. Furthermore, it is possible to have atomic but nondurable transactions.

**Supported operations.** Table VIII lists the operations that can be performed in a transaction. All systems that have transactions, except Redis PubSub, support both producing and consuming messages in a transaction. Redis PubSub is the only system that does not support message consumption in a transaction. In RabbitMQ, a consumer can receive a message and then decline to process it. RabbitMQ is the only system that allows rejecting the processing of messages and supports including the decision to reject a message as part of the transaction. Rejected messages are requeued to the queue, added to a dead-letter exchange, or deleted.

**Transaction scope.** Interestingly, as shown in Table VIII, all systems that support transactions allow transactions to access multiple topics. Therefore, an application can consume messages from one or more topics and produce messages to one or more topics in a single transaction.

### D. Message Ordering

Each MOM system provides different guarantees for the order in which messages are delivered to consumers. The ordering semantics define the order in which messages are delivered to consumers differently for two cases: for messages that are produced by the same producer and for messages produced by multiple producers to the same multicast exchange. We found that systems may offer two types of ordering guarantees, which we call *path ordering* and *multicast exchange ordering*. Table IX shows the message ordering semantics that each of the systems supports.

**Path ordering.** Path ordering means that if a producer sends two messages and these messages take the same path in the logical dissemination topology to the same consumer, then the order of these messages is preserved. Path ordering does not offer any ordering guarantees for multiple producers, even if the producers send messages to the same topic.

**Multicast exchange ordering.** Multicast exchange ordering offers stronger ordering guarantees than path ordering. Multicast exchange ordering guarantees that consumers of the same multicast exchange will receive all messages produced to that exchange in the same order regardless of which producer produced the messages or the path the messages took to the multicast exchange.

### E. Summary

The MOM systems we study adopt four message delivery semantics: best-effort at-most-once, reliable at-most-once, at-least-once, and effectively-once. Additionally, some systems

TABLE VIII
Transaction properties, supported operations, and scope.

| System | | Redis Streams | Redis PubSub | NSQ | Kafka | NATS | Pulsar | RabbitMQ | Ejabberd (eCS) | ActiveMQ Classic | ActiveMQ Artemis |
|---|---|---|---|---|---|---|---|---|---|---|---|
| Transaction Support | | ✓ | ✓ | | ✓ | | ✓ | ✓ | | ✓ | ✓ |
| Transaction Properties | Atomicity | | | | | | ✓ | | | ✓ | ✓ |
| | Isolation | ✓ | ✓ | | | | ✓ | ✓ | | ✓ | ✓ |
| Transaction Operations | Produce | ✓ | ✓ | | ✓ | | ✓ | ✓ | | ✓ | ✓ |
| | Consume & Acknowledge | ✓ | | | ✓ | | ✓ | ✓ | | ✓ | ✓ |
| | Reject | | | | | | | ✓ | | | |
| Transaction Scope | Single Topic | ✓ | ✓ | | ✓ | | ✓ | ✓ | | ✓ | ✓ |
| | Multiple Topics | ✓ | ✓ | | ✓ | | ✓ | ✓ | | ✓ | ✓ |

---

**Note 5: The design of Pulsar's transactional support.**

As an example, we detail the design of the transactional support in Pulsar. In Pulsar, a Transaction Coordinator (TC) module manages a transaction's life cycle. The TC stores the transaction state in three data structures:

1) *Transaction log*: A transaction log is a log stored on stable storage that stores a transaction's metadata. Pulsar implements the transaction log as a durable topic.

2) *Transaction buffers*: Each transaction has a buffer for each topic it uses to store messages produced within the transaction. Messages in a buffer are not visible to consumers until the transaction is committed.

3) *Pending Acknowledge State (PAS)*: It is a log that stores any acknowledgments sent during a transaction. When a transaction commits, all acknowledgments stored in the PAS are sent. The PAS is stored on durable storage. The TC also uses the PAS to detect conflicts between transactions. If a transaction acknowledges a message that already has a pending acknowledgment in the PAS log, the transaction will abort.

**Transaction Processing.** To start a transaction, a client contacts the TC. The TC logs the beginning of the transaction in the transaction log and assigns a unique ID to the transaction.

To produce a message, the client will first ask the TC to add the target topic to the transaction log. When the client produces messages, the broker stores the messages in a transaction buffer.

To acknowledge the consumption of a message, the client asks the TC to add the subscription to its log. The acknowledgments the client sends to a broker are stored in the PAS store.

To end a transaction, the client sends a request to the TC to commit or abort the transaction. If a transaction is aborted, the TC will log the decision, delete all transaction buffer(s), and delete its PAS store. If a transaction is committed, the TC logs the commit decision, moves messages from transaction buffer(s) to the actual topic(s) or topic(s) partition(s), and releases all acknowledgments stored in the PAS. Finally, the TC acknowledges the commit operation to the client. After this point, an offline garbage collection mechanism truncates the transaction log.

---

automatically detect and drop duplicate messages that result from message retransmission. These systems detect duplicate messages using an ID or a sequence number that the producer adds to each message upon production. The majority of MOM systems support carrying out operations within a transaction that accesses multiple topics. Transaction atomicity and isolation properties are not flexible, whereas the durability property is flexible and follows the durability configuration of the system. Lastly, in terms of the message ordering semantics that MOM systems support, there are two possible alternatives: path ordering and multicast exchange ordering. Multicast exchange ordering guarantees that all consumers of a multicast exchange will observe messages in the exact same order. Whereas, path ordering guarantees that messages of the same producer are delivered in order only if they take the same path in the message dissemination graph.

## VIII. Resource Management and Flow Control

In this section, we discuss the mechanisms implemented for resource management and flow control.

### A. Resource Management

We discuss the limits that MOM systems offer to control resource usage and the actions the systems take when these limits are violated. Table X shows the limits and violation policies. Table X uses symbols to match each resource limit to the violation policies applicable when the limit is violated. **Resource Limits.** Table X shows the resources that can be limited. We restrict our discussion to limits that are configurable (i.e., the user can set the resource limits). We note

TABLE IX
Supported message ordering semantics. J refers to a feature exclusively supported by NATS JetStream.

| System | | Redis Streams | Redis PubSub | NSQ | Kafka | NATS | Pulsar | RabbitMQ | Ejabberd (eCS) | ActiveMQ Classic | ActiveMQ Artemis |
|---|---|---|---|---|---|---|---|---|---|---|---|
| Message Ordering Guarantees | None | | ✓ | ✓ | | | | | | | |
| | Path ordering | ✓ | | | ✓ | ✓ | ✓ | ✓ | | ✓ | |
| | Multicast exchange | ✓ | | | ✓ | ✓J | ✓ | ✓ | | ✓[1] | ✓ |

(1) When the topology consists of a network of brokers, total ordering is not guaranteed.

TABLE X
The supported resource limits and the corresponding resource violation policies. The symbols match each resource limit to the violation policies applicable when the limit is violated. The network limit in NATS applies to both NATS Core and NATS JetStream.

| System | | Redis Streams | Redis PubSub | NSQ | Kafka | NATS JetStream | Pulsar | RabbitMQ | Ejabberd (eCS) | ActiveMQ Classic | ActiveMQ Artemis |
|---|---|---|---|---|---|---|---|---|---|---|---|
| Resource Limits | Rate Limit ($\varepsilon$) | | | | $\varepsilon$ | | | | | | |
| | Disk Space Limit ($\delta$) | | | | | $\delta$ | $\delta$ | $\delta$ | | $\delta$ | $\delta$ |
| | Memory Limit ($\mu$) | $\mu$ | $\mu$ | | | $\mu$ | | $\mu$ | $\mu$ | $\mu$ | $\mu$ |
| | Network Limit ($\kappa$) | $\kappa$ | $\kappa$ | $\kappa$ | $\kappa^1$ | $\kappa$ | $\kappa^2$ | $\kappa^3$ | $\kappa^3$ | $\kappa$ | $\kappa^3$ |
| | Topic Size Limit ($\sigma$) | $\sigma$ | | $\sigma$ | $\sigma$ | $\sigma$ | $\sigma$ | $\sigma$ | | $\sigma$ | $\sigma$ |
| Violation Policy | Spill to Disk | | | $\sigma$ | | | | $\mu\sigma$ | | $\mu\sigma$ | $\mu\sigma$ |
| | Block Client(s) | $\mu$ | $\mu$ | | $\varepsilon\kappa$ | $\delta\mu\sigma$ | $\delta$ | $\delta\mu\sigma$ | $\mu^4$ | $\delta\mu\sigma$ | $\delta\mu\sigma$ |
| | Notify | $\mu\kappa$ | $\mu\kappa$ | $\kappa$ | $\varepsilon\kappa$ | $\delta\mu\kappa\sigma$ | $\delta\kappa$ | $\delta\mu\kappa\sigma$ | $\kappa$ | $\delta\mu\kappa\sigma$ | $\delta\mu\kappa\sigma$ |
| | Refuse/Close Connection | $\kappa$ | $\kappa$ | $\kappa$ | $\kappa$ | $\kappa$ | $\kappa$ | $\kappa$ | $\kappa$ | | $\kappa$ |
| | Discard | $\mu\sigma$ | $\mu$ | $\sigma$ | $\sigma$ | $\sigma$ | $\sigma$ | $\sigma$ | | | $\sigma$ |

(1) Kafka allows this limit to be set per source IP.
(2) Configured using Websocket proxy or by setting separate limits on inbound and outbound connections.
(3) Connection limit applies per application.
(4) When it runs out of memory, Ejabberd can kill client-to-server and server-to-server connection sessions to reduce memory pressure.

that the following resource limits are per MOM broker (i.e., a violation of the limits by an application on one node does not affect the application on another node within the same MOM deployment).

- *Rate limit*: Only Kafka supports a rate limit (Table X). The Kafka configuration can be used to set a "request rate quota," which represents a percentage of the total processing and networking threads in the broker within a period of time. For instance, a 50% quota allows a client to use up to 50% of the Kafka broker threads in a window of time. This quota can be configured per client or for groups of clients and could be configured with different values on different brokers in the cluster.
- *Disk space limit*: This can be used to limit the amount of disk space that a broker uses. It can be specified as a percentage of the total disk space or set as an absolute value. NATS JetStream, Pulsar, RabbitMQ, ActiveMQ Classic, and ActiveMQ Artemis provide this configuration option.
- *Memory limit*: This configuration limits the total memory available for a broker. It can be specified as a percentage of the total memory or set as an absolute value. This limit is supported by Redis Streams, Redis PubSub, NATS JetStream, RabbitMQ, Ejabberd, ActiveMQ Classic, and ActiveMQ Artemis.
- *Network limit*: Table X shows that all the systems we study allow the number of active open connections to be limited.
- *Topic size limit*: This limit caps the number or total size of messages in a topic. This limit is offered by all the systems that we study except Redis PubSub, NATS Core, and Ejabberd.

**Violation Policies.** We found five violation policies (Table X) that are applicable when the aforementioned resource limits are reached.

- *Spill to disk*: When a topic size or a broker memory limit is reached, the system will spill messages to disk. This is supported by NSQ, RabbitMQ, ActiveMQ Classic, and ActiveMQ Artemis.
- *Block client(s)*: If a limit is reached, this policy blocks the offending client, a group of clients, or all clients. Table X shows that this policy is supported by nine of the systems we study.

- *Notify*: Following this policy, the broker sends a notification to producers or consumers that try to utilize a resource while its limit is reached. Table X shows that this is the most widely supported violation policy.
- *Refuse/close connection*: Table X shows that all systems, except Ejabberd, refuse new connections once the connection limit is reached. In Kafka, unless the limit is set per source IP, new connections are queued to wait for a vacancy. Otherwise, new connections will be immediately dropped. Ejabberd applies this policy per application and closes the oldest connection of an application before accepting a new connection from that application.
- *Discard*: This policy discards old messages to make room for new messages. The messages are discarded by removing them from the system. This policy is supported by Redis Streams, Redis PubSub, Kafka, NATS JetStream, Pulsar, RabbitMQ, and ActiveMQ Artemis. In NSQ, ephemeral topics discard old messages once the topic size limit is reached instead of spilling to disk. RabbitMQ also provides an option to move messages to a dead-letter exchange.

### B. Flow Control

Flow control is an essential technique to prevent brokers and consumers from being overwhelmed. Flow control mechanisms are applied between producers and brokers, and between brokers and consumers (or between producers and consumers for peer-to-peer systems, e.g., NSQ). The MOM systems that we study offer two types of flow control: credit-based and rate-based. Some of the systems that we study support both of these techniques and allow them to be used simultaneously. Table XI shows the flow control mechanism offered by each of the systems we study.

*1) Credit-Based Flow Control:* Table XI shows that several systems we study can apply a credit-based flow control, in which the number or size (i.e., total number of bytes) of buffered messages is limited. When the sender reaches this limit, it will stop sending messages until the receiver acknowledges some of the previous messages or grants additional credits. Noticeably, some of the systems that offer this flow control mechanism provide a configuration option to disable it.

**Producer-to-broker credit-based flow control.** Only ActiveMQ Classic and ActiveMQ Artemis support this type of flow control (Table XI). A producer requests credits from the broker. The broker responds with the number of credits allocated to that producer. The credits specify the total number of bytes of unacknowledged messages a producer is permitted. A producer can only send a message if it has available credit.

**MOM-system-to-consumer credit-based flow control.** To control the flow of messages from a broker, or a producer in the case of NSQ, to a consumer, MOM systems offer two options:

- *Consumer-specified credit*: In this approach, a consumer grants a broker credits specifying the number of messages it is ready to receive. When these credits are exhausted, the consumer sends a message to the broker with new credits or acknowledges a minimum number of messages. Redis Streams, NSQ, NATS JetStream, Pulsar, RabbitMQ, and ActiveMQ Classic support this option. In Ejabberd, a failed consumer can choose to poll missed messages upon recovery, where it can control the flow of messages by requesting a few messages at a time.
- *System-wide consumer preconfigured credit*: In this case, a broker has a preconfigured credit value for the number or total number of bytes of in-flight messages per consumer. NATS JetStream, Pulsar, RabbitMQ, Ejabberd, ActiveMQ Classic, and ActiveMQ Artemis support this approach.

*2) Rate-Based Flow Control:* Rate-based flow control limits the message rate or throughput of producers and consumers accessing a broker. Two options are available:

- *Producer rate limit*: Limits the throughput (messages or bytes per second) of a specific producer, group of producers, or every producer accessing a broker. Kafka offers a configuration option to limit data throughput per producer or group of producers. ActiveMQ Artemis limits the number of messages per producer upon the producer's request. RabbitMQ and ActiveMQ Classic automatically throttle fast producers for queues and consumers, respectively, to keep up with the pace.
- *Consumer rate limit*: Limits the throughput (messages or bytes per second) of a consumer, a group of consumers, or every consumer using the same broker. Kafka, NATS JetStream, Pulsar, and ActiveMQ Artemis support this rate limit. Kafka offers a configuration option to limit data throughput per consumer or group of consumers. Pulsar offers a configurable limit per subscription. Push-based consumers in NATS JetStream set their limits upon creation. Consumers in ActiveMQ Artemis inform the broker of their limit.

### C. Summary

In this section, we discuss the limits that MOM systems impose on the usage of available hardware resources along with the applicable violation policies. MOM systems can specify resource limits on processing, memory, network, and disk resources. To keep resource usage within these limits, MOM systems support five applicable violation policies, with notifying a client that tries to utilize an exhausted resource being the most supported policy. Additionally, point-to-point credit-based and rate-based flow control mechanisms are applicable to avoid overwhelming brokers and consumers, with some systems supporting both mechanisms. Finally, the resource management and flow control techniques are the base for offering isolation for multi-tenant deployments.

## IX. Active Messaging

In this section, we discuss the active messaging feature offered by Pulsar and Ejabberd. In this paradigm, messages can be processed using application-specific code while in transit through the MOM system.

These capabilities can be used to implement complex processing pipelines. Furthermore, active messaging can be

TABLE XI
THE SUPPORTED FLOW CONTROL MECHANISMS.

| System | | Redis Streams | Redis PubSub | NSQ | Kafka | NATS JetStream | Pulsar | RabbitMQ | Ejabberd (eCS) | ActiveMQ Classic | ActiveMQ Artemis |
|---|---|---|---|---|---|---|---|---|---|---|---|
| Credit-Based Flow Control | Producer-Requested | | | | | | | | | ✓ | ✓ |
| | Consumer-Specified | ✓ | | ✓ | | ✓ | ✓ | ✓ | ✓[3] | ✓ | |
| | Consumer-Preconfigured | | | | | ✓[1] | ✓[2] | ✓ | ✓[4] | | |
| Rate-Based Flow Control | Producer Rate | | | | | ✓ | | ✓ | ✓ | ✓ | ✓ |
| | Consumer Rate | | | | ✓ | ✓[5] | ✓[6] | | | | ✓ |

(1) NATS JetStream offers a flow control mechanism for consumers that share a subscription. It uses a preconfigured sliding-window flow control approach. It cannot be configured but can be disabled.
(2) A preconfigured limit can be set on shared subscriptions.
(3) With client flexible offline message retrieval.
(4) Applies when active stream management is utilized, where the client session closes once the consumer accumulates a configurable number of unacknowledged messages.
(5) Configured upon the creation of push-based consumers.
(6) Can be configured per subscription.

used to deploy a flexible dissemination topology and support custom load balancing and filtering techniques.

**Pulsar Active Messaging**. Active messaging in Pulsar is enabled through *Pulsar Functions*, which are serverless computing functions written in Java, Python, or Go. Pulsar Functions can consume messages from one or more input topics, apply user-defined processing logic, and produce messages to one or more output topics. Pulsar runs the functions on the broker or on a dedicated computational cluster. Pulsar Functions support chaining, in which the output topic of a function can be the input topic of another function. Pulsar Functions feature three types of processing semantics:

- *At-most-once:* A message is immediately acknowledged after being consumed by the function, regardless of whether or not the message is processed successfully.
- *At-least-once:* A message is guaranteed to be processed at least once. However, a message can be processed multiple times under some failure scenarios.
- *Effectively-once:* Similar to the at-least-once semantic, a message can be consumed multiple times by a function, but the output of the function will be produced once in the output topics. This is done by logging the function's status using durable storage and checking the function status before committing the output of the function.

**Ejabberd Active Messaging**. Ejabberd leverages its modular system design and Erlang hot code swapping to enable active messaging. The Ejabberd architecture relies on minimal system core modules, whereas the majority of features are developed as pluggable modules. This modular architecture allows the system to be extended with new modules to provide new features. In addition, Ejabberd allows the system administrator to load new modules, disable a running module, or enable a module during runtime.

## X. CLIENT INTERACTION

In this section, we discuss the mechanisms through which a client interacts with the system. We detail configuration settings related to how clients discover services and produce and consume messages.

### A. Discovery Services

A discovery service helps producers and consumers find the information needed to access the system. At a minimum, this information is a list of brokers in the system. Some systems also help clients find the exact broker that serves a given topic. Furthermore, the discovery service can also help with masking broker failures by pointing producers and consumers to alternative brokers.

Table XII shows the discovery services and their consistency semantics for the systems we studied. All of the systems that we study offer a built-in discovery service. RabbitMQ and ActiveMQ Classic can also use an external discovery service. RabbitMQ can be configured to use the AWS (EC2) [101] discovery service, Kubernetes [102], Consul [103], and etcd [104]. ActiveMQ Classic supports the use of the Lightweight Directory Access Protocol (LDAP) [105] server.

**Discovery service consistency semantics**. The consistency semantics of the discovery service specify when the service reflects a change in the system state. We found that the discovery services in the systems we study have either strong consistency guarantees or eventual consistency guarantees (Table XII):

- *Strongly consistent*: The state of the system is stored on a replicated and linearizable discovery service. This consistency guarantee is offered by Kafka, Pulsar, and RabbitMQ. Kafka and Pulsar use ZooKeeper, which offers a strongly consistent discovery service. RabbitMQ supports plugins for a discovery service. Among the supported plugins are Consul and etcd, which are based on the Raft consensus algorithm [99].

TABLE XII
SUPPORTED DISCOVERY SERVICES AND THEIR CONSISTENCY SEMANTICS.

| System | | Redis Streams | Redis PubSub | NSQ | Kafka | NATS | Pulsar | RabbitMQ | Ejabberd (eCS) | ActiveMQ Classic | ActiveMQ Artemis |
|---|---|---|---|---|---|---|---|---|---|---|---|
| Discovery Service | Built-in Service | ✓ | ✓ | ✓ | ✓ | ✓ | ✓ | ✓ | ✓ | ✓ | ✓ |
| | Add-On Service | | | | | | | ✓ | ✓ | ✓ | |
| Consistency of Discovery Service | Eventually Consistent | ✓ | ✓ | ✓ | | ✓ | | ✓[1] | ✓ | ✓ | ✓ |
| | Strongly Consistent | | | | ✓ | | ✓ | ✓ | | | |

(1) RabbitMQ can use an external system to provide a discovery service including Kubernetes. Kubernetes semantics can be relaxed to offer eventual consistency.

> **Note 6: Discovery service design**
> Some of the systems we study document the design of their discovery services. Here, we present an overview of the design techniques used in these systems.
> **Strongly consistent discovery service.** Kafka, Pulsar, and RabbitMQ provide strongly consistent discovery services that use a linearizable coordination service such as Zookeeper or RAFT-based storage to maintain the system metadata.
> **Eventually consistent discovery service.** MOM systems with an eventually consistent discovery service adopted one of two designs. Redis Streams, Redis PubSub, and ActiveMQ Classic can be configured to use an eventually consistent storage service to build a discovery service. Redis Stream and Redis PubSub use Redis Sentinel for their discovery service. Redis Sentinel is a replicated and eventually consistent coordination service. ActiveMQ Classic has a similar option in which it can use an external eventually consistent LDAP service [105].
> NATS, ActiveMQ Classic, and ActiveMQ Artemis adopt a peer-to-peer approach to build a discovery service. The configuration file includes a list of seed brokers that clients and new brokers can contact to access the service. Nodes in the cluster periodically multicast information about newly joined or departed nodes. NATS uses a gossip-based approach to disseminate updates. In ActiveMQ Artemis, every broker creates a connection to every other broker in the system and uses these connections to send periodic updates. Alternatively, ActiveMQ Artemis brokers can multicast their updates using IP multicasting or a JGroup reliable all-to-all messaging service [106].

- *Eventually consistent*: In this consistency model, the discovery service might serve stale results about the state of the system, such as providing the address of a broker that is no longer a member of the system, or returning the address of an older broker for a certain topic. Table XII shows that all systems except Kafka and Pulsar offer an eventually consistent discovery service.

### B. Access Methods

The systems we study offer the following approaches for consumers to obtain messages from the system. Table XIII shows the access methods supported by each of these systems.

- *Push*: The broker, or the producer in the case of the peer-to-peer NSQ system, pushes messages to consumers, then each consumer processes the messages it has received through an event listener. We find that all systems we study except Redis Streams and Kafka offer a push-based approach (Table XIII). Furthermore, this is typically the default access mechanism in systems that support multiple access mechanisms. We note that this is the only access mechanism offered by Redis PubSub, NSQ, and NATS Core. This approach offers low latency; therefore, it is often used in latency-sensitive applications. Pushing messages may overwhelm consumers and lead to long delays or loss of messages. Consequently, systems that offer this approach often provide a flow control mechanism to create back pressure when the consumer is overloaded. We discuss the supported flow control mechanisms in Section VIII-B.

- *Pull*: The consumer requests one or more messages from the broker and blocks until it receives the messages. Redis Streams, Kafka, and ActiveMQ Classic support this approach.
- *Poll*: The consumer requests one or more messages from the broker, but does not block indefinitely if no messages are available. This approach is supported by Redis Streams, Kafka, NATS JetStream, RabbitMQ, Ejabberd, ActiveMQ Classic, and ActiveMQ Artemis.
- *Fetch*: The consumer can selectively request previously processed messages. A message can be identified with a unique ID or an index in the log. Table XIII shows that Redis Streams, Kafka, NATS JetStream, Pulsar, Ejabberd, ActiveMQ Classic, and ActiveMQ Artemis offer this access mechanism. Pulsar offers special read-only consumers that are the only ones that can fetch older messages. In Ejabberd, fetch is the only access method

TABLE XIII

The supported consumer access methods for message consumption. Each access method has a unique symbol that is used to correlate the access method with its granularity. J refers to a feature exclusively supported by NATS JetStream.

| System | | Redis Streams | Redis PubSub | NSQ | Kafka | NATS | Pulsar | RabbitMQ | Ejabberd (eCS) | ActiveMQ Classic | ActiveMQ Artemis |
|---|---|---|---|---|---|---|---|---|---|---|---|
| Consumer Access Method | Push ($\rho$) | | $\rho$ | $\rho$ | | $\rho$[1] | $\rho$ | $\rho$ | $\rho$ | $\rho$ | $\rho$ |
| | Pull ($\mu$) | $\mu$ | | | $\mu$ | | | | | $\mu$ | |
| | Poll (o) | o | | | o | oJ | | o | o[4] | o | o |
| | Fetch (↻) | ↻ | | | ↻ | ↻J | ↻[3] | | ↻[4] | ↻[5] | ↻[5] |
| | Notify ($\nu$) | | | | | | | | $\nu$ | | |
| Configuration Granularity | Request | $\mu$ o ↻ | | | $\mu$ o ↻ | o[2] ↻ | | | | [$\mu$o][6] | |
| | Consumer | | | | | $\rho$ o[2] | ↻[3] | $\rho$o | o[4] | $\rho$[$\mu$o][6]↻ | $\rho$o ↻ |
| | System | | $\rho$ | $\rho$ | | $\rho$[1] | $\rho$ | | $\rho$ ↻ $\nu$ | | |

(1) The Push access method is the only option in NATS Core.
(2) After declaring a JetStream consumer to be poll-based, it sends poll and fetch requests.
(3) A Pulsar consumer using the reader interface can fetch specific messages.
(4) Fetch applies only for uploaded files and archived messages. Poll applies only if the recovered consumer chooses flexible offline message retrieval over flooding.
(5) Can be done through creating a consumer that uses filtering to retrieve specific messages.
(6) A consumer can be created to be push-based or poll/pull-based. Poll/pull-based consumers poll/pull on a per-request basis.

for uploaded files and archived messages.

- *Notify*: A notification of a new message is pushed to the consumer, but the actual message is not forwarded until the consumer requests the message. Ejabberd is the only system that supports this approach for some modules.

**Access Method Configuration Granularity**. Table XIII shows the three granularities at which access methods are controlled:

- *Request*: A consumer specifies the access method in every request to retrieve messages. The same consumer can use different access methods for different requests. This is the only granularity available in Redis Streams and Kafka.
- *Consumer*: A consumer configuration specifies the access method. NATS JetStream, Pulsar, RabbitMQ, Ejabberd, ActiveMQ Classic, and ActiveMQ Artemis support this approach. NATS JetStream offers the ability to configure consumers as push-based or poll-based consumers. NATS JetStream poll-based consumers can issue poll or fetch requests. A request-based consumer in ActiveMQ Classic can poll/pull messages depending on the request timeout parameter. In Ejabberd, recovered consumers can choose to poll missed messages to avoid being flooded if these messages were pushed by the server.
- *System*: A system-wide configuration can control the access method used for all consumers and messages. Redis PubSub, NSQ, and NATS Core only support this configuration granularity. In Pulsar, push is the default system-wide access mechanism unless the consumer uses the reader interface to consume specific messages. In Ejabberd, push is the default system-wide access mechanism unless the system is configured to omit the inclusion of the message payload inside the notification. In addition, fetch is the only supported access mechanism if message archiving is enabled.

### C. Dissemination Policies for the Unicast Exchange

All the systems we study, except Redis PubSub and Ejabberd, support both multicasting and unicasting of messages. For systems that support unicast exchanges, if an exchange has multiple consumers, the next message will be sent to one of them. Table XIV shows the policies used to select a consumer in this case.

- *Random*: The broker, or the producer in the case of the NSQ peer-to-peer system, selects a destination to receive the next message, or a subset of messages, using a uniform random distribution. NSQ and NATS readily support this policy. In Kafka, messages for a topic are divided into partitions and each partition can be assigned to a consumer. The default assignment of messages to partitions is done in a uniform random fashion.
- *Round-Robin*: The broker selects a destination to receive the next message, or subset of messages, in a round-robin fashion. Pulsar, RabbitMQ, and ActiveMQ support this policy.
- *FCFS*: The next consumer to request or indicate that it is ready to receive message(s) will receive the message(s). This policy is often used by request-based consumers in Redis Streams, NATS JetStream, and RabbitMQ. Alternatively, disabling the buffering of messages on the consumer side for push-based consumers in Pulsar, ActiveMQ Classic, and ActiveMQ Artemis achieves this policy.
- *Failover*: The broker will pick one consumer to receive all messages sent to the unicast exchange. Should that consumer fail, the broker will pick another consumer to receive the messages. The selected consumer can be a

TABLE XIV

THE SUPPORTED DISSEMINATION POLICIES PER SYSTEM. J REFERS TO A FEATURE EXCLUSIVELY SUPPORTED BY NATS JETSTREAM.

| System | | Redis Streams | Redis PubSub | NSQ | Kafka | NATS | Pulsar | RabbitMQ | Ejabberd (eCS) | ActiveMQ Classic | ActiveMQ Artemis |
|---|---|---|---|---|---|---|---|---|---|---|---|
| Multicast | | ✓ | ✓ | ✓ | ✓ | ✓ | ✓ | ✓ | ✓ | ✓ | ✓ |
| Unicast | | ✓ | | ✓ | ✓ | ✓ | ✓ | ✓ | | ✓ | ✓ |
| Load Balancing Policy | Random | | | ✓ | ✓ | ✓ | | | | | |
| | Round-Robin | | | | | | ✓ | ✓ | | ✓ | ✓ |
| | First-Come-First-Served (FCFS) | ✓ | | | | ✓J[1] | ✓[2] | ✓[1] | | ✓[2] | ✓[2] |
| | Failover | | | | | | ✓ | ✓ | | ✓ | ✓ |
| | Consumer Priority | | | | | | ✓ | ✓ | | ✓ | ✓ |
| | Affinity-Based | | | | ✓ | | ✓ | | | ✓ | ✓ |

(1) Applies only to poll-based consumers.
(2) For push-based consumers, FCFS can be achieved by disabling buffering on the consumer side. This way, each consumer will consume one message at a time which achieves FCFS semantics.

consumer that subscribed to the same unicast exchange or a consumer sharing a multicast exchange subscription with the failed consumer. Pulsar, RabbitMQ, and ActiveMQ support this policy.

- *Priority*: Pulsar, RabbitMQ, ActiveMQ Classic, and ActiveMQ Artemis allow clients to assign priorities to consumers. Consumers buffer messages for processing. When a consumer's buffer is full, the broker stops sending messages to this consumer. When the consumer acknowledges some of the messages, the broker sends more messages to this consumer. When consumers have priority levels, the broker sends messages to the consumers with the highest priority until their buffers are full. Then, the broker selects the lower priority consumers. If multiple consumers with the same priority level exist, then the messages will be distributed using the default policy among the high-priority consumers. In all of the aforementioned systems, the default policy is round-robin. Section VIII-B discusses in detail the consumer side buffering and flow control mechanism.

- *Affinity based*: A MOM system may allow messages to be grouped based on an ID, a key, or a hash of a subset of header fields. Following the affinity-based policy, the broker forwards messages that are part of the same group to the same consumer.

A use case that specifically benefits from a variety of policies is scheduling. In this use case, producers send requests that are load balanced to multiple consumers (or servers in this scenario). Random, round-robin, and FCFS policies try to utilize all consumers. Failover aims to use a single consumer to ensure in-order processing of messages; once that consumer fails, another consumer is selected to replace it. An affinity-based policy can be used to send all messages produced by a producer, or share a grouping, to the same server.

The priority policy can be used to implement auto-scaling of cloud services. The messaging system sends all requests to a few high-priority servers. If the system receives too many requests, the system deploys additional low-priority servers and sends the additional requests to them. If the load decreases, the system shuts down low-priority servers.

### D. Multi-Tenancy

Some of the systems we study support multi-tenancy, which means that the system allows multiple applications to use the same MOM system instance. An essential aspect of supporting multi-tenancy is providing the illusion that the system is not shared with other applications. This requires isolating the applications so that one application does not affect the performance of others.

Kafka, NATS, Pulsar, RabbitMQ, and Ejabberd support multi-tenancy. These systems offer isolation between different tenant's configuration, data, and performance. Each tenant has its own topics, dissemination topology, and data. Each tenant also has its own configuration and authentication setup. Most importantly, these systems provide performance isolation through resource quotas and rate limits.

### E. Message Content

In this section, we describe our observations about the content of messages, including the payload and the metadata attributes carried in a message. Table XV shows the content and metadata types supported by the systems we study.

**Message Payload.** Table XV shows that the systems we study support three options for message payloads:

- *Opaque (Byte Stream)*: All of the systems we study, except Redis Streams and Kafka, support sending opaque messages that do not follow a specific data format.
- *Key-Value Pair*: Redis Streams and Kafka support sending the message payload as a key-value pair. A message may contain multiple key-value pairs.
- *Object*: ActiveMQ Classic and ActiveMQ Artemis support setting the message payload as an object as part of their support of the Java Message Service (JMS) standard. Nevertheless, the ActiveMQ documentation recommends

TABLE XV
THE DIFFERENT TYPES OF MESSAGE CONTENTS AND ATTRIBUTES WE OBSERVE IN THE TEN STUDIED SYSTEMS.

| System | | Redis Streams | Redis PubSub | NSQ | Kafka | NATS | Pulsar | RabbitMQ | Ejabberd (eCS) | ActiveMQ Classic | ActiveMQ Artemis |
|---|---|---|---|---|---|---|---|---|---|---|---|
| Message Content Type | Opaque | | ✓ | ✓ | | ✓ | ✓ | ✓ | ✓ | ✓ | ✓ |
| | Key-Value Pair(s) | ✓ | | | ✓ | | | | | | |
| | Object | | | | | | | | | ✓ | ✓ |
| Message Headers | MIME-Type | | | | ✓ | | | ✓ | | ✓ | ✓ |
| | Map and List (Key-Value Sequence) | | | | | | ✓ | | | ✓ | ✓ |
| | Message Priority | | | | | | | ✓ | | ✓ | ✓ |

TABLE XVI
THE SUPPORTED FILTERING MECHANISM.

| System | | Redis Streams | Redis PubSub | NSQ | Kafka | NATS | Pulsar | RabbitMQ | Ejabberd (eCS) | ActiveMQ Classic | ActiveMQ Artemis |
|---|---|---|---|---|---|---|---|---|---|---|---|
| Filtering | Topic | ✓ | ✓ | ✓ | ✓ | ✓ | ✓ | ✓ | ✓ | ✓ | ✓ |
| | Headers | | | | | | | ✓ | | ✓ | ✓ |
| | Custom Content-Based | | | | | | ✓[1] | | ✓[2] | | |

(1) Can be implemented using Pulsar Functions.
(2) Can be implemented using Ejabberd custom modules.

against its use because it introduces coupling between producers and consumers.

**Message Attribute.** Table XV shows that the systems we study support three options for message attributes:

- *MIME-Type Header*: Kafka, RabbitMQ, ActiveMQ Classic, and ActiveMQ Artemis allow adding a media-type identifier (MIME-type) to the message payload.
- *Map and List (Key-Value Sequence)*: Pulsar, ActiveMQ Classic, and ActiveMQ Artemis allow the attribute format to be a sequence of key-value pairs. The key-value attributes can be application defined.
- *Message Priority*: RabbitMQ, ActiveMQ Classic, and ActiveMQ Artemis allow the producer to set a priority value for a message. Messages with higher priority are delivered before messages with lower priority. Consumer applications can also read the priority level of a message and use it to prioritize message processing or to filter messages based on their priority level. We noticed that in systems that support message priorities, there is no limit on the number of messages that a producer can designate as high-priority messages.

### F. Message Filtering

MOM systems can filter messages and selectively forward them to a subset of the consumers. Table XVI lists the filtering options available in the systems we study:

- *Topic*: This is the standard filtering technique for all MOM systems. Producers produce messages to a topic and the messages are forwarded to consumers subscribed to the topic. To distribute the load among brokers, Kafka divides messages of a topic into partitions and assigns partitions to different brokers. The placement of messages in partitions is based on the hashing of a key provided by the producer. In this way, a producer can ensure that related messages are placed in the same partition. If the producer does not provide a key, messages will be randomly assigned to partitions. We note that the partitioning function can be overridden and replaced with a custom version. Pulsar can also divide messages of a topic into partitions.
- *Headers*: This strategy filters messages based on one or more fields in the message header. This filtering is supported by RabbitMQ, ActiveMQ Classic, and ActiveMQ Artemis (Table XVI). ActiveMQ Classic and ActiveMQ Artemis offer an SQL-based selector on the header fields to select messages.
- *Custom Content-Based*: Pulsar and Ejabberd support application-specific custom filtering. Pulsar allows application developers to implement a custom filtering function to decide how messages are divided among the partitions.

### G. Message Acknowledgment

The MOM systems that we study may support message acknowledgment in their interactions with producers and consumers. A broker can acknowledge messages to producers, and consumers can acknowledge messages to a broker (or a producer in the NSQ peer-to-peer system). An acknowl-

TABLE XVII

The supported message acknowledgment modes. NSQ does not offer acknowledgments to producers and Redis PubSub does not support consumer acknowledgments.

| System | | Redis Streams | Redis PubSub | NSQ | Kafka | NATS JetStream | Pulsar | RabbitMQ | Ejabberd (eCS) | ActiveMQ Classic | ActiveMQ Artemis |
|---|---|---|---|---|---|---|---|---|---|---|---|
| Producer Confirms | Controlled | | | | | ✓² | | ✓ | ✓⁴ | ✓⁵ | |
| Confirmation Mode | Single | ✓ | ✓ | | | ✓² | ✓ | | | ✓ | ✓ |
| | Cumulative | | | | | | | ✓ | ✓⁴ | | |
| | Multiple | | | | | | | | | | |
| | Batching | | | | ✓ | | ✓³ | | | | |
| | Transactional | | | | | | ✓ | | | ✓ | ✓ |
| Consumer Acknowledgments | Controlled | ✓¹ | | | | | ✓ | ✓ | ✓⁴ | ✓ | ✓ |
| Acknowledgment Mode | Single | ✓¹ | | ✓ | | | ✓ | ✓ | | ✓ | ✓ |
| | Cumulative | | | | ✓ | ✓ | ✓ | ✓ | ✓⁴ | ✓ | ✓ |
| | Multiple | ✓¹ | | | | | | | | ✓ | |
| | Batching | | | | | | ✓³ | | | | |
| | Transactional | | | | | | ✓ | | | ✓ | ✓ |

(1) Applies only to consumer groups. For regular consumers, no acknowledgment is supported.
(2) Enabled only when the producer chooses to produce messages with the special JetStream Publish command and messages are then confirmed individually by default.
(3) Messages that are produced as a batch must be acknowledged as a batch.
(4) Supported only when the stream management module is enabled.
(5) It depends on the durability of the produced message. Non-persistent messages are not confirmed; however, persistent messages must be confirmed.

edgment sent to the producer confirms that the broker has received the message, and in some cases has stored it on disk (e.g., ActiveMQ Classic and ActiveMQ Artemis) or replicated it on backup brokers (e.g., Kafka), but it typically does not indicate that a consumer has received the message. A consumer acknowledgment may indicate that the consumer has successfully received, processed, or stored the message.

The systems we study support a range of options for delivering these acknowledgments. Table XVII lists the options that each system supports. The following options apply per topic (i.e., one can acknowledge multiple messages from the same topic, not across topics). Table XVII shows that all systems other than NSQ support producer-side acknowledgment and all systems except Redis PubSub and NATS Core support consumer-side acknowledgment.

- *Single*: Acknowledgments are sent per message. This is the most widely supported option. In NSQ, this is the only supported consumer acknowledgment option and it limits per-consumer throughput because the consumer has to send an acknowledgment for each message.
- *Cumulative*: Acknowledging a single message ID implicitly acknowledges all previous messages from the producer or broker with smaller message IDs. All systems, except Redis and NSO, support this acknowledgment mode.
- *Multiple*: Multiple, not-necessarily consecutive, messages are acknowledged in a single acknowledgment reply. This mode is supported by Redis Streams for consumers that share a consumer group and by ActiveMQ Classic Opti-mized Acknowledgment Mode, where acknowledgments for a range of messages can be combined in a single operation.
- *Batch*: A group of messages is produced or consumed as a single unit and one acknowledgment is sent per batch. This is supported by Kafka and Pulsar on the producer side and only by Pulsar on the consumer side.
- *Transaction*: In this mechanism, messages are acknowledged as part of an atomic transaction. Pulsar, ActiveMQ Classic, and ActiveMQ Artemis offer this option on the producer and consumer sides. We discuss support for transactions in Section VII-C.

Table XVII shows that most systems offer one or two acknowledgment modes, whereas Pulsar, ActiveMQ Classic, and ActiveMQ Artemis offer a wide range of acknowledgment options.

### H. Message Discard Policies

Messages are generally discarded upon consumption, but they can be retained until some conditions are met. Redis Streams, NATS JetStream, and Ejabberd allow messages to be retained indefinitely. In this section, we discuss message discard policies. We do not discuss discarding messages due to resource limitations or flow control violations in this section, and we refer the reader to Section VIII for a detailed discussion of these issues.

We found that four message discard policies are supported. Messages that satisfy the conditions of those policies can be deleted or moved to a special topic called "dead-letter

TABLE XVIII
The supported retention policies, their granularity, and ways messages are discarded once a retention policy is violated. Each retention policy has a unique symbol that is used to show the granularity and the discard actions related to that policy.

| | System | Redis Streams | Redis PubSub | NSQ | Kafka | NATS JetStream | Pulsar | RabbitMQ | Ejabberd (eCS) | ActiveMQ Classic | ActiveMQ Artemis |
|---|---|---|---|---|---|---|---|---|---|---|---|
| Discard Policy | Acknowledgment ($\alpha$) | | | $\alpha$ | | $\alpha$ | $\alpha$ | $\alpha$ | $\alpha^4$ | $\alpha$ | $\alpha$ |
| | Reject/Remove ($\varepsilon$) | $\varepsilon$ | | | | $\varepsilon$ | $\varepsilon$ | $\varepsilon$ | $\varepsilon$ | | |
| | Delivery Count ($\mu$) | $\mu$ | | | | | $\mu$ | $\mu^2$ | | $\mu$ | $\mu$ |
| | Time-To-Live ($\tau$) | | | | $\tau$ | $\tau$ | $\tau$ | $\tau$ | $\tau$ | $\tau$ | $\tau$ |
| Policy Granularity | Per Message | $\varepsilon$ | | | | $\varepsilon$ | | $\varepsilon^3\,\tau$ | $\varepsilon$ | $\tau$ | $\varepsilon\,\tau$ |
| | Per Topic | | | | $\tau$ | $\alpha\,\tau$ | | $\mu\,\tau$ | $\tau^5$ | $\mu$ | $\mu$ |
| | Per Subscription | | | | | | $\varepsilon^1\mu$ | | | | |
| | Per Consumer | $\mu$ | | | | | | $\alpha\,\varepsilon^3$ | $\alpha^4$ | $\alpha$ | |
| | System-wide/Default | | | $\alpha$ | | | $\alpha\,\tau$ | | $\alpha^4\tau^6$ | | $\alpha$ |
| Discard Actions | Completely Removed | $\varepsilon$ | | $\alpha$ | $\tau$ | $\alpha\,\varepsilon\,\tau$ | $\alpha\,\varepsilon\,\tau$ | $\alpha\,\varepsilon\,\mu\,\tau$ | $\alpha\,\varepsilon\,\tau$ | $\alpha\,\tau$ | $\alpha\,\varepsilon\,\tau$ |
| | Dead-letter | $\mu$ | | | | | $\mu$ | $\varepsilon\,\mu\,\tau$ | | $\mu$ | $\mu\,\tau$ |

(1) Pulsar allows the removal of old messages in a subscription (unicast exchange, in our model) that are older than a given time.
(2) Delivery count is only supported when using the replicated and durable "quorum queues."
(3) In RabbitMQ, message rejection is done per message but only by certain consumers that have the "manual acknowledgment" configuration option enabled.
(4) Supported only when active stream management is enabled between the communicating parties.
(5) Applies to PubSub topics. If a per topic value is not defined, then the system default applies.
(6) Applies to the uploaded files.

exchange." These policies can be controlled through configuration options or API calls at different granularities. Table XVIII shows the policies, granularities, and actions supported by the systems we study. In the table, each policy has a unique symbol. We use the same symbol to show the granularity and the discard action associated with each policy. Redis PubSub and NATS Core are the only systems that do not have a configurable message discard mechanism.

**Discard Policy.** The discard policy sets the conditions for discarding messages. Table XVIII shows the discard policies supported by each of the systems we study.

- *Acknowledgment*: A message is deleted once acknowledged by the targeted consumer, or consumers, subscribed to that topic. This is the default policy in most MOM systems, it is supported by seven of the ten systems we study. This discard policy always results in the removal of the acknowledged messages. Pulsar further allows a time to be set to keep acknowledged messages in the system. For instance, an application can choose to keep the processed messages for 24 hours after processing for auditing or debugging purposes.

- *Rejection and explicit removal*: A message is discarded when a consumer rejects the processing of it, or a producer or consumer requests that a message be deleted. This policy usually leads to deleting the selected message.

- *Delivery counter*: If the broker does not receive an acknowledgment from a consumer, then the broker will try to redeliver the message. A "delivery counter" counts the number of times the broker tried to deliver the message. This policy will discard a message if the counter exceeds a configurable threshold. Redis Streams, Pulsar, RabbitMQ, ActiveMQ Classic, and ActiveMQ Artemis support this policy.

- *Time To Live (TTL)*: A message is discarded once its TTL timer expires. TTL is defined as a period (i.e., seconds, minutes, or even days). Kafka, NATS JetStream, Pulsar, RabbitMQ, Ejabberd, ActiveMQ Classic, and ActiveMQ Artemis support this policy.

**Discard Actions.** If a message meets the condition of one of the policies discussed above, then one of two actions will be taken:

- *The message will be removed from the system*: The exact message that violates the policy will be deleted from the system. This action is supported by all the systems that support discarding messages.

- *The message will be moved to a dead-letter exchange*: A message meeting a discard condition is moved to a special topic called dead-letter exchange. An application can choose to inspect messages in the dead-letter exchange (e.g., for debugging purposes).

**Discard Policy Granularity.** The systems we study offer multiple granularities at which the discard policy is controlled.

- *Per message*: The discard policy is controlled per message. A consumer or producer can request the removal of a specific message or messages, or a producer can set per-message attributes to control the discard policy. For instance, in RabbitMQ, ActiveMQ Classic, and ActiveMQ Artemis, a producer can specify a TTL value for a produced message. Once that TTL expires, the message is deleted or moved to a dead-letter exchange.

- *Per topic*: The discard policy is controlled per topic. For instance, a per-topic configuration can specify the TTL for each message produced to that topic in Kafka, NATS JetStream, and Ejabberd's PubSub module. Additionally, the system configuration in RabbitMQ, ActiveMQ Classic, and ActiveMQ Artemis can specify the delivery counter for a topic. If a message in that topic exceeds this counter, it is deleted or moved to a dead-letter exchange.
- *Per subscription*: Pulsar offers a configuration option to set a delivery count policy per subscription. A subscription in Pulsar is equivalent to a unicast exchange in our model (Section IV) in which multiple consumers can share a subscription.
- *Per consumer*: The discard policy is controlled by consumers. For instance, a message in Redis Streams contains its delivery count, and then it is up to the consumer to decide whether to process it or move it to a dead-letter exchange.
- *System-wide/default*: The discard policy is either applied as the default policy to all messages in the system or is controlled through a system-wide configuration.

Table XVIII shows some interesting patterns: The TTL policy is typically applied per message or topic. Messages that exceed the delivery count are typically moved into a dead-letter exchange. Acknowledged and explicitly removed messages are typically deleted from the system. Explicit removals are always issued per message except in the case of Pulsar, in which a prefix of the message log can be discarded in one request. Finally, we note that some systems allow the use of multiple discard policies simultaneously. If a message satisfies the conditions of one of the enabled policies, the message is discarded.

### I. Summary

Clients in all systems retrieve the information they need to access the system through a discovery service. Discovery services typically offer eventually consistent semantics. Then, consumers receive messages from the MOM system using one of five approaches, with pushing messages to consumers being the most common. When multiple consumers subscribe to a unicast exchange, the exchange load balances the messages across the consumers using one of six dissemination policies, where the majority of systems supports multiple alternatives. Lastly, some systems support true multi-tenancy by isolating tenants' data, configuration, and resource usage.

### XI. Messaging Protocols

Several protocols have been developed that specify how clients can interact with MOM systems, including JMS, AMQP, STOMP, MQTT, XMPP, and OpenWire. These protocols typically define specifications for interacting with a MOM system, help integrate and interoperate multiple MOM systems, remove vendor lock-in, and provide a common software layer that supports popular standards. These protocols provide specifications at different levels, from low-level message formats to delivery semantics, to application-level API. For example, JMS offers a high-level Java-based API that defines how a client can interact with a MOM system. All of the protocols mentioned above provide wire-level specifications (i.e., the structure and format of messages). Furthermore, all of the listed protocols specify some of the semantics for the MOM system, such as the access mechanisms, and transactions or delivery semantics.

Table XIX shows the protocols supported by the systems we study. Seven systems use custom protocols, five of which only use a custom protocol. MQTT is the most popular standard and is supported by five of the systems we study. We note that RabbitMQ, ActiveMQ Classic, and ActiveMQ Artemis support five or more protocols.

### XII. Implementation Details

MOM systems often offer a client library that allows applications to access the MOM system. In this section, we discuss implementation details related to client interactions.

#### A. Transport Protocols

Table XX shows the transport layer protocols and communication modes (blocking/non-blocking) of the systems we study. All MOM systems support the use of TCP to transfer messages. Noticeably, Pulsar, Ejabberd, and ActiveMQ Classic offer the option of using UDP to transfer messages. Six of the systems we study offer WebSocket-based communication. Finally, ActiveMQ Classic is the only system that can leverage IP multicasting to deliver messages to consumers.

#### B. MOM-Consumer Communication Modes

We now examine the mode of communication (blocking or non-blocking) that is initiated from the MOM service to the consumer. The mode of communication from producers to the MOM systems depends on the producer design, so it is outside the scope of the implementation of the MOM system.

Table XX shows the modes of communication offered by the systems we study. We note that all the systems support a non-blocking mode of communication in their implementation. In this mode, a MOM service dispatches messages to a consumer without blocking to wait for an acknowledgment. This design avoids blocking on slow consumers and reduces the number of threads used for message forwarding. We note that ActiveMQ Classic is the only system that allows the choice of whether the mode of communication should be blocking or non-blocking. Moreover, ActiveMQ Classic allows this configuration to be set per consumer.

#### C. HTTP Interface

Typically, MOM systems offer a Command Line Interface (CLI) to manage and configure the system. Some MOM systems offer an HTTP interface that allows users to interact with the system. The HTTP interface allows four operation types:

- Producing and consuming messages.
- Creating new topics.
- Changing system configurations.

TABLE XIX

THE MESSAGING STANDARDS SUPPORTED BY THE SYSTEMS WE STUDY. P MEANS A STANDARD IS SUPPORTED THROUGH A PLUG-IN. J MEANS THAT THE STANDARD IS EXCLUSIVELY SUPPORTED IN NAT JETSTREAM.

| System | | Redis Streams | Redis PubSub | NSQ | Kafka | NATS | Pulsar | RabbitMQ | Ejabberd (eCS) | ActiveMQ Classic | ActiveMQ Artemis |
|---|---|---|---|---|---|---|---|---|---|---|---|
| Messaging Protocols | JMS | | | | | | | ✓P | | ✓ | ✓ |
| | AMQP 0.9.1 | | | | | | | ✓ | | | |
| | AMQP 1.0 | | | | | | | ✓P | | ✓ | ✓P |
| | STOMP | | | | | | | ✓P | | ✓ | ✓P |
| | MQTT | | | | | ✓J | | ✓P | ✓P | ✓ | ✓P |
| | XMPP | | | | | | | | ✓ | ✓ | |
| | OpenWire | | | | | | | | | ✓ | ✓P |
| | Other Standard | | | | | | | | | ✓ | ✓P |
| | Custom | ✓ | ✓ | ✓ | ✓ | ✓ | ✓ | | | | ✓ |

TABLE XX

THE SUPPORTED TRANSPORT PROTOCOLS AND MOM-CONSUMER COMMUNICATION MODES.

| System | | Redis Streams | Redis PubSub | NSQ | Kafka | NATS | Pulsar | RabbitMQ | Ejabberd (eCS) | ActiveMQ Classic | ActiveMQ Artemis |
|---|---|---|---|---|---|---|---|---|---|---|---|
| Transport Protocols | TCP | ✓ | ✓ | ✓ | ✓ | ✓ | ✓ | ✓ | ✓ | ✓ | ✓ |
| | UDP | | | | | | | ✓ | ✓ | ✓ | |
| | WebSocket | | | | | ✓ | ✓ | ✓[1] | ✓ | ✓ | ✓[2] |
| | IP Multicast | | | | | | | | | ✓ | |
| MOM-Consumer Communication Mode | Non-blocking | ✓ | ✓ | ✓ | ✓ | ✓ | ✓ | ✓ | ✓ | | ✓ |
| | Blocking | | | | | | | | | ✓ | |

(1) Only supported by the STOMP and MQTT plugins.
(2) ActiveMQ Artemis supports Websocket transport for the AMQP, STOMP, and MQTT protocols.

TABLE XXI

THE OPERATIONS SUPPORTED THROUGH THE HTTP INTERFACE.

| System | | Redis Streams | Redis PubSub | NSQ | Kafka | NATS | Pulsar | RabbitMQ | Ejabberd (eCS) | ActiveMQ Classic | ActiveMQ Artemis |
|---|---|---|---|---|---|---|---|---|---|---|---|
| HTTP Interface | Produce\Consume | | | ✓ | | | | ✓ | ✓ | ✓ | ✓ |
| | Create Topics | | | ✓ | | | ✓ | ✓ | ✓ | ✓ | ✓ |
| | Configuration | | | ✓ | | | ✓ | ✓ | ✓[1] | | ✓ |
| | Metrics | | | ✓ | | | ✓ | ✓ | ✓ | ✓ | ✓ |

(1) Configuration changes through the HTTP interface only affect the current running instance and are not permanent. If the system reboots, then the configuration is lost.

TABLE XXII
The security measures supported in the ten systems we studied.

| System | | Redis Streams | Redis PubSub | NSQ | Kafka | NATS | Pulsar | RabbitMQ | Ejabberd (eCS) | ActiveMQ Classic | ActiveMQ Artemis |
|---|---|---|---|---|---|---|---|---|---|---|---|
| Security Measures | Built-in Authentication | ✓ | ✓ | ✓ | ✓ | ✓ | ✓ | ✓ | ✓ | ✓ | ✓ |
| | External Authentication | | | ✓ | ✓ | | ✓ | ✓ | ✓ | ✓ | ✓ |
| | Access Control List (ACL) | ✓ | ✓ | | ✓ | ✓ | ✓ | ✓ | ✓ | ✓ | ✓ |
| | Secure Transport Layer | ✓ | ✓ | ✓ | ✓ | ✓ | ✓ | ✓ | ✓ | ✓ | ✓ |

• Monitoring the state of the system and its performance.

Table XXI shows the systems that offer an HTTP interface and the supported operations. NSQ, Pulsar, RabbitMQ, Ejabberd, ActiveMQ Classic, and ActiveMQ Artemis offer an HTTP interface. Table XXI shows that NSQ, RabbitMQ, Ejabberd, and ActiveMQ Artemis support all the aforementioned types of operations. Pulsar does not support producing or consuming messages through the HTTP interface. ActiveMQ Classic does not support changing the system configuration through the HTTP interface. Pulsar and ActiveMQ Classic offer a RESTful API.

### D. Security

Table XXII lists the supported security measures to provide secure client interactions. The majority of the systems we study offer standard security mechanisms for authentication, access control, and secure communication.

• *Authentication*: All of the ten systems that we study support a range of authentication options, including a username/password, certificate, or access token for each client. Moreover, many of the systems support using an external authentication system, making it easier to integrate them into an institution's IT infrastructure.

• *Access Control List (ACL)*: All the systems we study, except NSQ, provide the ACL mechanism. ACL specifies what a client can access and with which operations (e.g., produce, subscribe, consume, or manage topics such as create, delete, and configure).

• *Secure communication*: All the systems we study support using a secure transport layer such as TLS [107] or SSL [108].

### E. Summary

In this section, we study the implementation details related to client interactions. All MOM systems that we study support TCP, whereas some systems also support UDP, Websocket, and IP Multicast as a message transport protocol. Additionally, all of the systems support non-blocking I/O in their implementation. Some MOM systems offer an HTTP interface to facilitate client access to the system in addition to the client libraries they typically offer. Lastly, the MOM systems that we study support standard mechanisms for client authentication, access control, and secure communication.

## XIII. Concluding Remarks

We conducted an in-depth analysis of ten popular MOM systems. For each system, we studied 42 features related to various aspects of MOM systems, including dissemination topologies, reliability guarantees, service semantics, client interaction mechanisms, resource management, flow control, and active messaging. We present our findings and identify open research problems. Our annotated data set is publicly available at [30] to help researchers understand the state-of-the-art systems, help developers choose the best system for a certain application, help practitioners understand the semantics of different systems, and help the community understand the capabilities of different systems and focus its effort on a fewer number of MOM systems.

**Insights and Research Challenges**. MOM systems incorporate features central to cloud applications and deployments, such as space and time decoupling, flexibility, and extensibility. Many of the MOM systems we study are designed with extreme flexibility in mind, starting from a flexible dissemination topology to supporting all possible configuration options and service semantics. This flexibility comes at a cost in system complexity, development, and testing.

Our data set [30] shows that systems like Redis PubSub, NATS Core, and NSQ support the least amount of features compared to others. For NSQ, this is the result of adopting a peer-to-peer brokerless topology that does not offer most of the MOM communication characteristics including space decoupling, time decoupling, and flexibility. Furthermore, the brokerless topology limits the number of features that can be supported, such as replication, subscription recovery, and message ordering per multicast exchange due to the lack of coordination between producers of a certain topic. On the other hand, Redis PubSub and NATS Core do not support features such as durable storage, replication, subscription recovery, detection of duplicate messages, consumer acknowledgments, message retention, and flow control, because they are designed to be best-effort in-memory MOM systems.

The data set [30] also shows that Pulsar, ActiveMQ Classic, and ActiveMQ Artemis support most of the features that we study. Comparing ActiveMQ distributions to pulsar reveals that ActiveMQ does not support rack-aware replica placement, multicast exchange ordering, a discovery service with strong consistency semantics, or content-based filtering. ActiveMQ distributions lack the active messaging feature that Pulsar offers, which allows one to implement flexible topologies, custom content-based filtering, and other application-specific processing logic. On the other hand, Pulsar lacks the support of brokerless peer-to-peer topology and flexible topology, object messages, message priority, limiting memory utilization, support of standard messaging protocols, and message transfer over IP multicast.

A large number of systems offer basic transaction support, with many offering non-atomic transactions. Supporting atomic and durable transactions for low-latency and high-throughput MOM systems remains an open research problem.

Active messaging is emerging as a new capability in MOM systems. This technology is in its infancy and has limited support for protecting a user from competing tenants or misbehaving functions. Building efficient resource management and isolation techniques for active messaging remains an open research problem.

## REFERENCES

[1] V. Venkataramani, Z. Amsden, N. Bronson et al., "TAO: How Facebook Serves the Social Graph," in Proceedings of the 2012 ACM SIGMOD International Conference on Management of Data, ser. SIGMOD '12. New York, NY, USA: Association for Computing Machinery, 2012, pp. 791–792. [Online]. Available: https://doi.org/10.1145/2213836.2213957

[2] B. Calder, J. Wang, A. Ogus et al., "Windows Azure Storage: A Highly Available Cloud Storage Service with Strong Consistency," in Proceedings of the Twenty-Third ACM Symposium on Operating Systems Principles, ser. SOSP '11. New York, NY, USA: Association for Computing Machinery, 2011, pp. 143–157. [Online]. Available: https://doi.org/10.1145/2043556.2043571

[3] S. Ghemawat, H. Gobioff, and S.-T. Leung, "The Google File System," in Proceedings of the 19th ACM Symposium on Operating Systems Principles, Bolton Landing, NY, 2003, pp. 20–43.

[4] "Hermes," http://hermes.allegro.tech/#why, accessed: October, 2021.

[5] "Building a Microservices Ecosystem with Kafka Streams and KSQL," https://www.confluent.io/blog/building-a-microservices-ecosystem-with-kafka-streams-and-ksql/, accessed: October, 2021.

[6] "Microservices - why use RabbitMQ? - CloudAMQP," https://www.cloudamqp.com/blog/why-use-rabbitmq-in-a-microservice-architecture.html, accessed: October, 2021.

[7] "Apache Storm," https://storm.apache.org/, accessed: October, 2021.

[8] "KAFKA STREAMS," https://kafka.apache.org/documentation/streams/, accessed: October, 2021.

[9] "An Introduction to Stream Processing with Pulsar Functions - DZone," https://dzone.com/articles/an-introduction-to-stream-processing-with-pulsar-f, accessed: October, 2021.

[10] "Is Kafka the Next Big Thing in the Banking and Financial Sector? - DZone," https://dzone.com/articles/is-kafka-the-next-big-thing-in-the-banking-and-fin, accessed: October, 2021.

[11] "How Kafka Helped Rabobank Modernize Alerting System," https://www.datanami.com/2017/08/15/kafka-helped-rabobank-modernize-alerting-system/, accessed: October, 2021.

[12] A. Al-Fuqaha, M. Guizani, M. Mohammadi et al., "Internet of Things: A Survey on Enabling Technologies, Protocols, and Applications," IEEE Communications Surveys Tutorials, vol. 17, no. 4, pp. 2347–2376, 2015.

[13] L. Rodríguez-Gil, J. García-Zubia, Pablo Orduña et al., "An Open and Scalable Web-Based Interactive Live-Streaming architecture: The WILSP Platform," IEEE Access, vol. 5, pp. 9842–9856, 2017.

[14] F. Wang, D. Zhang, Y. Lu et al., ""PSVA: A Content-Based Publish/Subscribe Video Advertising Framework"," in Smart Computing and Communication, M. Qiu, Ed. Cham: Springer International Publishing, 2018, pp. 249–258.

[15] H. Hoang, B. Cassell, T. Brecht et al., "RocketBufs: A Framework for Building Efficient, In-Memory, Message-Oriented Middleware," in Proceedings of the 14th ACM International Conference on Distributed and Event-Based Systems, ser. DEBS '20. New York, NY, USA: Association for Computing Machinery, 2020, pp. 121–132. [Online]. Available: https://doi.org/10.1145/3401025.3401744

[16] "Using Apache Kafka to Drive Cutting-Edge Machine Learning — Confluent," https://www.confluent.io/blog/using-apache-kafka-drive-cutting-edge-machine-learning, accessed: October, 2021.

[17] A. Khlif, "A distributed computation system for deep learning experiments with Docker Compose and RabbitMQ." https://deezer.io/a-distributed-computation-system-for-deep-learning-experiments-with-docker-compose-and-rabbitmq-5ac4ab344406, accessed: October, 2021.

[18] Z. Han and M. Xu, "Machine Learning Techniques in Storm," in 2015 Seventh International Symposium on Parallel Architectures, Algorithms and Programming, Nanjing, China, 2015, pp. 139–142.

[19] E. Androulaki, A. Barger, V. Bortnikov et al., "Hyperledger Fabric: A Distributed Operating System for Permissioned Blockchains," in 12th USENIX Symposium on Networked Systems Design and Implementation (NSDI 15), Oakland, CA, 2018, pp. 351–366.

[20] "Apache Kafka," https://kafka.apache.org/, accessed: October, 2021.

[21] "Hello from Apache Pulsar — Apache Pulsar," https://pulsar.apache.org/, accessed: October, 2021.

[22] "Messaging that just works - RabbitMQ," https://www.rabbitmq.com/, accessed: October, 2021.

[23] "ActiveMQ Artemis The Next Generation Message Broker by ActiveMQ." https://activemq.apache.org/components/artemis/, accessed: October, 2021.

[24] "MQ - IBM MQ - Canada — IBM," https://www.ibm.com/ca-en/products/mq, accessed: October, 2021.

[25] "IronMQ - Serverless Message Queue," http://www.iron.io/mq, accessed: October, 2021.

[26] Y. Sharma, P. Ajoux, P. Ang et al., "Wormhole: Reliable Pub-Sub to Support Geo-Replicated Internet Services," in Proceedings of the 12th USENIX Conference on Networked Systems Design and Implementation, ser. NSDI'15. USA: USENIX Association, 2015, pp. 351–366.

[27] "Google Cloud Pub/Sub," https://cloud.google.com/pubsub, accessed: October, 2021.

[28] "Fully Managed Message Queuing - Amazon Simple Queue Service - Amazon Web Services," https://aws.amazon.com/sqs/, accessed: October, 2021.

[29] "Azure Queue storage documentation — Microsoft Learn," https://docs.microsoft.com/en-us/azure/storage/queues/, accessed: October, 2021.

[30] "A Survey of Message-Oriented Middleware Systems," https://docs.google.com/spreadsheets/d/1HrZ7ub19FuuBzA5z4aA6RfR5vnkdnm0bg3hxfADspEA/edit?usp=sharing, accessed: October, 2021.

[31] P. T. Eugster, P. A. Felber, R. Guerraoui et al., "The Many Faces of Publish/Subscribe," ACM Comput. Surv., vol. 35, no. 2, pp. 114–131, Jun. 2003. [Online]. Available: http://doi.acm.org.proxy.lib.uwaterloo.ca/10.1145/857076.857078

[32] "TIBCO Rendezvous®," https://www.tibco.com/products/tibco-rendezvous, accessed: October, 2021.

[33] G. Banavar, T. Chandra, R. Strom, and D. Sturman, ""A Case for Message Oriented Middleware"," in Distributed Computing, P. Jayanti, Ed. Berlin, Heidelberg: Springer Berlin Heidelberg, 1999, pp. 1–17.

[34] J. Yongguo, L. Qiang, Q. Changshuai, S. Jian, and L. Qianqian, "Message-oriented middleware: A review," in 2019 5th International Conference on Big Data Computing and Communications (BIGCOM), 2019, pp. 88–97.

[35] T. R. Sheltami, A. A. Al-Roubaiey, and A. S. Mahmoud, "A Survey on Developing Publish/Subscribe Middleware over Wireless Sensor/Actuator Networks," Wirel. Netw., vol. 22, no. 6, pp. 2049–2070, Aug 2016. [Online]. Available: https://doi.org/10.1007/s11276-015-1075-0

[36] E. Souto, G. Guimares, G. Vasconcelos et al., "Mires: A publish/subscribe middleware for sensor networks," Personal Ubiquitous Comput., vol. 10, pp. 37–44, 02 2006.

[37] J.-H. Hauer, V. Handziski, A. Köpke et al., "A Component Framework for Content-Based Publish/Subscribe in Sensor Networks," in Wireless Sensor Networks. Berlin, Heidelberg: Springer Berlin Heidelberg, 2008, pp. 369–385.

[38] K. Shi, Z. Deng, and X. Qin, "TinyMQ: A Content-based Publish/Subscribe Middleware for Wireless Sensor Networks," in The international conference on sensor technologies and application, 2011.

[39] Y. Liu, L.-J. Zhang, and C. Xing, "Review for message-oriented middleware," in Internet of Things - ICIOT 2020: 5th International Conference, Held as Part of the Services Conference Federation, SCF 2020, Honolulu, HI, USA, September 18-20, 2020, Proceedings. Berlin, Heidelberg: Springer-Verlag, 2020, p. 152159. [Online]. Available: https://doi.org/10.1007/978-3-030-59615-6_12

[40] M. A. Razzaque, M. Milojevic-Jevric, A. Palade et al., "Middleware for Internet of Things: A Survey," IEEE Internet of Things Journal, vol. 3, no. 1, pp. 70–95, 2016.

[41] K. Sachs, S. Kounev, J. Bacon et al., "Performance Evaluation of Message-Oriented Middleware Using the SPECjms2007 Benchmark," Performance Evaluation, vol. 66, pp. 410–434, 08 2009.

[42] K. Sachs, S. Appel, S. Kounev et al., "Benchmarking Publish/Subscribe-Based Messaging Systems," in Proceedings of the 15th International Conference on Database Systems for Advanced Applications, ser. DASFAA'10. Berlin, Heidelberg: Springer-Verlag, 2010, pp. 203–214.

[43] "OpenMessaging benchmark," https://openmessaging.cloud/, accessed: October, 2021.

[44] A. Ahuja, V. Jain, and D. Saini, Characterization and Benchmarking of Message-Oriented Middleware. Cham: Springer International Publishing, 2021, pp. 129–147. [Online]. Available: https://doi.org/10.1007/978-3-030-75614-7_9

[45] V. Jain, A. Ahuja, and D. Saini, "Evaluation and Performance Analysis of Apache Pulsar and NATS," in Cyber Security and Digital Forensics, K. Khanna, V. V. Estrela, and J. J. P. C. Rodrigues, Eds. Singapore: Springer Singapore, 2022, pp. 179–190.

[46] V. John and X. Liu, "A Survey of Distributed Message Broker Queues," CoRR, vol. abs/1704.00411, 2017. [Online]. Available: http://arxiv.org/abs/1704.00411

[47] "GitHub - vineetjohn/flotilla: Automated message queue orchestration for scaled-up benchmarking." https://github.com/vineetjohn/flotilla, accessed: October, 2021.

[48] "Redis Streams," https://redis.io/docs/data-types/streams/, accessed: October, 2021.

[49] "Redis Enterprise Software - The Real-Time Data Platform," https://redis.com/redis-enterprise-software/overview/, accessed: October, 2021.

[50] "Redis Pub/Sub," https://redis.io/docs/manual/pubsub/, accessed: October, 2021.

[51] "NSQ A realtime distributed messaging platform," https://nsq.io/, accessed: October, 2021.

[52] "Data in Motion Platform for Enterprise," https://www.confluent.io/product/confluent-platform/, accessed: October, 2021.

[53] "Apache RocketMQ," https://rocketmq.apache.org/, accessed: October, 2021.

[54] "What is Message Queue for Apache RocketMQ?" https://www.alibabacloud.com/help/doc-detail/29532.htm?spm=a2c63.l28256.a3.10.48687882EsHNSE, accessed: October, 2021.

[55] "NATS.io - Cloud Native, Open Source, High-performance Messaging," https://nats.io/, accessed: October, 2021.

[56] "Synadia," https://synadia.com/, accessed: October, 2021.

[57] "StreamNative BYOC: Pulsar-as-a-Service in the cloud of your choice," https://streamnative.io/cloud/managed/, accessed: October, 2021.

[58] "VMware RabbitMQ Documentation," https://docs.vmware.com/en/VMware-Tanzu-RabbitMQ/index.html, accessed: October, 2021.

[59] "EMQX: The World's #1 Open Source Distributed MQTT Broker," https://www.emqx.io/, accessed: October, 2021.

[60] "EMQX Enterprise: Enterprise MQTT Platform At Scale," https://www.emqx.com/en/products/emqx?utm_source=emqx.io&utm_medium=referral&utm_campaign=emqxio-header-to-enterprise, accessed: October, 2021.

[61] "Eclipse Mosquitto," https://mosquitto.org/, accessed: October, 2021.

[62] "TIBCO® Messaging - Eclipse Mosquitto Distribution," https://www.tibco.com/products/tibco-messaging-eclipse-mosquitto-distribution, accessed: October, 2021.

[63] "ejabberd XMPP Server with MQTT Broker & SIP Service," https://www.ejabberd.im/, accessed: October, 2021.

[64] "Create Awesome Realtime Systems XMPP Server + MQTT Broker + SIP Service," https://www.process-one.net/en/ejabberd/#getejabberd, accessed: October, 2021.

[65] "Faye: Simple pub/sub messaging for the web," https://faye.jcoglan.com/, accessed: October, 2021.

[66] "Emitter: Scalable Real-Time Communication Across Devices," https://emitter.io/, accessed: October, 2021.

[67] "GitHub - emitter-io/emitter: High performance, distributed and low latency publish-subscribe platform." https://github.com/emitter-io/emitter#licensing, accessed: October, 2021.

[68] "Nchan - flexible pubsub for the modern web," https://nchan.io/, accessed: October, 2021.

[69] "VerneMQ MQTT broker," https://vernemq.com/, accessed: October, 2021.

[70] "Commercial Services - VerneMQ," https://vernemq.com/services.html, accessed: October, 2021.

[71] "ActiveMQ "Classic" The Tried and Trusted Open Source Message Broker," https://activemq.apache.org/components/classic/, accessed: October, 2021.

[72] "ActiveMQ," https://activemq.apache.org/support#commercial-support-, accessed: October, 2021.

[73] "Github - moscajs/aedes: Barebone mqtt broker that can run on any stream server, the node way," https://github.com/moscajs/aedes, accessed: October, 2021.

[74] "Aedes," https://github.com/moscajs/aedes#support, accessed: October, 2021.

[75] "GitHub - hivemq/hivemq-community-edition: HiveMQ CE is a Java-based open source MQTT broker that fully supports MQTT 3.x and MQTT 5. It is the foundation of the HiveMQ Enterprise Connectivity and Messaging Platform," https://github.com/hivemq/hivemq-community-edition, accessed: October, 2021.

[76] "Discover the 3 different editions of HiveMQ," https://www.hivemq.com/hivemq/editions/, accessed: October, 2021.

[77] "Federation Plugin - RabbitMQ," https://www.rabbitmq.com/federation.html, accessed: October, 2021.

[78] "Shovel Plugin - RabbitMQ," https://www.rabbitmq.com/shovel.html, accessed: October, 2021.

[79] "Virtual Destinations," https://activemq.apache.org/virtual-destinations, accessed: October, 2021.

[80] "Address Federation. ActiveMQ Artemis Documentation," https://activemq.apache.org/components/artemis/documentation/latest/federation-address.html, accessed: October, 2021.

[81] "GitHub REST API - GitHub Docs," https://docs.github.com/en/rest, accessed: October, 2021.

[82] H. Borges, A. Hora, and M. T. Valente, "Understanding the Factors That Impact the Popularity of GitHub Repositories," in 2016 IEEE International Conference on Software Maintenance and Evolution (ICSME), 2016, pp. 334–344.

[83] M. Nagappan, T. Zimmermann, and C. Bird, "Diversity in Software Engineering Research," in Proceedings of the 2013 9th Joint Meeting on Foundations of Software Engineering, ser. ESEC/FSE 2013. New York, NY, USA: Association for Computing Machinery, 2013, pp. 466–476. [Online]. Available: https://doi.org/10.1145/2491411.2491415

[84] "ZeroMQ," https://zeromq.org/, accessed: October, 2021.

[85] "GitHub - nanomsg/nanomsg: nanomsg library," https://github.com/nanomsg/nanomsg, accessed: October, 2021.

[86] "The WhatsApp Architecture Facebook Bought For $19 Billion - High Scalability," http://highscalability.com/blog/2014/2/26/the-whatsapp-architecture-facebook-bought-for-19-billion.html, accessed: October, 2021.

[87] "Ubisoft — ProcessOne," https://www.process-one.net/en/customers/ubisoft/, accessed: October, 2021.

[88] "How LinkedIn customizes Apache Kafka for 7 trillion messages per day — LinkedIn Engineering," https://engineering.linkedin.com/blog/2019/apache-kafka-trillion-messages, accessed: October, 2021.

[89] "Can Kafka Handle a Lyft Ride? - Confluent," https://www.confluent.io/resources/kafka-summit-2020/can-kafka-handle-a-lyft-ride/, accessed: October, 2021.

[90] "Scaling with RabbitMQ @ Soundcloud," https://tanzu.vmware.com/content/blog/scaling-with-rabbitmq-soundcloud, accessed: October, 2021.

[91] "Open-sourcing Pulsar, Pub-sub Messaging at Scale — Yahoo Engineering," https://yahooeng.tumblr.com/post/150078336821/open-sourcing-pulsar-pub-sub-messaging-at-scale#notes?ref_url=https://yahooeng.tumblr.com/post/150078336821/open-sourcing-pulsar-pub-sub-messaging-at-scale/embed#_=_, accessed: October, 2021.

[92] "rpc(3) - Linux manual page," https://man7.org/linux/man-pages/man3/rpc.3.html, accessed: October, 2021.

[93] "gRPC," https://grpc.io/, accessed: October, 2021.

[94] "Apache Thrift - Home," https://thrift.apache.org/, accessed: October, 2021.

[95] "Rebtel — ProcessOne," https://www.process-one.net/en/customers/rebtel/, accessed: October, 2021.

[96] "Nimbuzz — ProcessOne," https://www.process-one.net/en/customers/nimbuzz/, accessed: October, 2021.

[97] "Hello from Apache BookKeeper — Apache BookKeeper," https://bookkeeper.apache.org/, accessed: October, 2021.

[98] "Mnesia," https://elixirschool.com/en/lessons/storage/mnesia, accessed: October, 2021.

[99] D. Ongaro and J. Ousterhout, "In Search of an Understandable Consensus Algorithm," in 2014 USENIX Annual Technical Conference (USENIX ATC 14). Philadelphia, PA: USENIX Association, Jun. 2014, pp. 305–319. [Online]. Available: https://www.usenix.org/conference/atc14/technical-sessions/presentation/ongaro

[100] A. Silberschatz, H. F. Korth, and S. Sudarshan, Database System Concepts, Seventh Edition. McGraw-Hill Book Company, 2020. [Online]. Available: https://www.db-book.com/db7/index.html

[101] "Amazon EC2 Web Services," https://aws.amazon.com/ec2/, accessed: October, 2021.

[102] "Kubernetes," https://kubernetes.io/, accessed: October, 2021.

[103] "Consul by HashiCorp," https://www.consul.io/, accessed: October, 2021.

[104] "etcd," https://etcd.io/, accessed: October, 2021.

[105] "LDAP.com - Lightweight Directory Access Protocol," https://ldap.com/, accessed: October, 2021.

[106] "JGroups - The JGroups Project," http://www.jgroups.org/, accessed: October, 2021.

[107] "Transport Layer Security - Wikipedia," https://en.wikipedia.org/wiki/Transport_Layer_Security, accessed: October, 2021.

[108] A. Weaver, "Secure Sockets Layer," Computer, vol. 39, no. 4, pp. 88–90, 2006.

## APPENDIX A: KEYWORDS

In this appendix, we list the unique keywords we use to search GitHub for MOM systems. We used these keywords and combinations of them. The exact list of search terms we use is available in our data set.

- AMQP
- ActiveMQ
- Atom
- Auto
- Broker
- Bus
- Framework
- HTTP
- HornetQ
- JMS
- Kafka
- MQTT
- Message
- Messaging
- Middleware
- Notification
- OpenWire
- Platform
- Pub
- Pub-Sub
- PubSub
- Publish
- Publish-subscribe
- Publish/Subscribe
- PublishSubscribe
- Publisher
- PublisherSubscriber
- Qpid
- Queue
- REST
- RSS
- Redis
- RocketMQ
- STOMP
- Server
- Storm
- Stream
- Streaming
- Sub
- Subscribe
- Subscriber
- System
- WSIF
- WS Notification
- XMPP